\def\draftversion{false}

\RequirePackage{ifthen}
\ifthenelse{\equal{\draftversion}{true}}{
  \documentclass[aps,prl,galley,showpacs,preprintnumbers,citeautoscript,
      amsmath,amssymb,longbibliography]{revtex4-1}
}{
  \documentclass[citeautoscript,floatfix,aps,prl,twocolumn,
      superscriptaddress,longbibliography]{revtex4-1}
}

\usepackage{amsmath}
\usepackage{graphicx}
\usepackage{epstopdf}
\usepackage{natbib}
\usepackage{array}
\usepackage{dcolumn}
\usepackage{bm}
\usepackage{multirow}
\usepackage{soul}  

\usepackage[usenames,dvipsnames]{color}

\newcommand\T{\rule{0pt}{2.6ex}}              
\newcommand\B{\rule[-1.2ex]{0pt}{0pt}}        

\soulregister\cite7

\ifthenelse{\equal{\draftversion}{true}}{
  \usepackage{showlabels}
  \marginparwidth 2.7in
  \marginparsep 0.5in
  \newcounter{comm} 
  \def\commnext{\stepcounter{comm}}
  \def\commtext{{\bf\color{blue}[\arabic{comm}]}}
  \def\commmar{{\bf\color{blue}[\arabic{comm}]}}
  \def\dvm#1{\commnext\marginpar{\small DV\commmar: #1}\commtext}
  \def\cdm#1{\commnext\marginpar{\small CED\commmar: #1}\commtext}
  \def\msm#1{\commnext\marginpar{\small MS\commmar: #1}\commtext}
  \def\asm#1{\commnext\marginpar{\small AS\commmar: #1}\commtext}
  \def\miq#1{\commnext\marginpar{\small MR\commmar: #1}\commtext}
  \def\mlab#1{\marginpar{\small\bf #1}}
  
}{
  \def\dvm#1{}
  \def\cdm#1{}
  \def\msm#1{}
  \def\asm#1{}
  \def\miq#1{}
  \def\mlab#1{}
  
}

\begin{document}

\title{In-plane flexoelectricity in two-dimensional $D_{3d}$ crystals}

\author{Matteo Springolo}
\affiliation{Institut de Ci\`encia de Materials de Barcelona 
(ICMAB-CSIC), Campus UAB, 08193 Bellaterra, Spain}

\author{Miquel Royo}
\email{mroyo@icmab.es}
\affiliation{Institut de Ci\`encia de Materials de Barcelona 
(ICMAB-CSIC), Campus UAB, 08193 Bellaterra, Spain}

\author{Massimiliano Stengel}
\email{mstengel@icmab.es}
\affiliation{Institut de Ci\`encia de Materials de Barcelona 
(ICMAB-CSIC), Campus UAB, 08193 Bellaterra, Spain}
\affiliation{ICREA - Instituci\'o Catalana de Recerca i Estudis Avan\c{c}ats, 08010 Barcelona, Spain}

\date{\today}

\begin{abstract} 

We predict a large in-plane polarization response to  
bending in a broad class of trigonal two-dimensional crystals.
We define and compute the relevant flexoelectric coefficients from first principles 
as linear-response properties of the undistorted layer, by using the primitive crystal 
cell.
The ensuing response (evaluated for SnS$_{2}$, silicene, phosphorene and RhI$_{3}$ monolayers and for a hexagonal BN bilayer) 
is up to one order of magnitude larger than the out-of-plane components in the same material. 
We illustrate the topological implications of our findings by calculating 
the polarization textures that are associated with a variety of rippled and bent structures.
We also determine the longitudinal electric fields induced by a flexural phonon at leading
order in amplitude and momentum.
\end{abstract}

\maketitle

The electromechanical properties of two-dimensional (2D) materials have generated considerable research interest recently, due to their potential application in flexible, stretchable, foldable, and wearable nanoscale devices~\cite{Fan20164283,Peng2023,Pandey2023}.
Materials design paradigms have traditionally relied on 
piezoelectricity, whose in-plane components are active in a broad 
variety of hexagonal monolayers~\cite{wu-14,zhu2014,Fei2015,dong2017}.  
More recently flexoelectricity (the generation of a macroscopic polarization or voltage in response to a strain gradient)
has emerged as an appealing higher-order alternative, thanks to its favorable scaling with decreasing sample size~\cite{nguyen2013}. 
Indeed, 2D materials can sustain large bending deformations without fracturing,~\cite{akinwande2017} which in principle
should enable a large flexoelectric response.

Flexoelectricity in 2D materials has been mostly studied as an out-of-plane dipolar response due to a flexural (bending) deformation.
Several studies of graphene, BN, phosphorene, transition-metal dichalcogenides (TMDs) and related materials have reported a great 
variety of magnitudes and signs for the relevant flexoelectric coefficient, sometimes with significant mutual disagreement.
On the experimental side~\cite{brennan-17,brennan-20,mcgilly-20,HAQUE202169,SUN2022107701,Hirakata_2022}, such inconsistencies 
likely originate from the difficulty in extracting the intrinsic response of the 2D monolayer from the measured signal. 
The scatter in the theoretical values, on the other hand, can be either traced back to
the broad variety of methods that were used in each case, 
ranging from first principles techniques~\cite{YakobsonCPL02,kalinin-08,shi-18,shi-19,pandey-21a,pandey-21b,codony2021,kumar-21} to classical~\cite{zhuang-19,javvaji-19,Yang-22}
and machine-learning~\cite{JAVVAJI2021558,Javvaji-22} force fields;  
or to the technical subtleties in defining the dipole moment of a curved surface. 
These difficulties can now be largely avoided via a publicly available~\cite{PhysRevX.9.021050,romero-20,Royo-22} linear-response 
implementation~\cite{springolo2021direct}, which is aimed at providing reference first-principles values for
the direct and converse flexoelectric effect in arbitrary 2D systems.

While the aforementioned out-of-plane response is universal, it turns out to be small in many materials.
In the present Letter, we investigate the \emph{in-plane} polarization response to a flexural deformation instead,
which we indicate as ``unconventional'' as it is typically ruled out in 3D crystals of sufficiently high symmetry.
The effect, illustrated in Figs.~1(a) and~(b), is \emph{linear} in the layer curvature, and therefore differs from 
the nonlinear one predicted in monolayer h-BN~\cite{naumov2009unusual} some time ago. It is active in a 
surprisingly broad class of 2D systems, including many of the best studied materials.
Interestingly, for materials with $D_{3d}$ point symmetry (space groups P$\bar{3}m1$ and P$\bar{3}1m$), the amplitude 
of the polarization response is insensitive to the bending direction, while its orientation continuously rotates 
in plane, with an angular periodicity that matches the three-fold axis of the flat configuration. [Figs.~1(b) and~(c)]
Apart from the obvious practical interest for energy-harvesting applications~\cite{duerloo2013flexural}, 
such an effect is important for fundamental reasons as well.
Most notably, it leads to a broad range of topologically nontrivial polarization textures in 
rippled and bent geometries, including vortices/antivortices and spontaneously polarized tubes.
%
Moreover, it endows flexural phonons with longitudinal electric fields (and hence with a long-ranged 
contribution to the electron-phonon Hamiltonian), which may significantly contribute to the mobility in the 
low-temperature regime.~\cite{Fischetti2016,Gunst2017,ponce2022drift,ponce2022long}

\begin{figure}[b!]
\begin{center}
 \includegraphics[width=1.0\columnwidth]{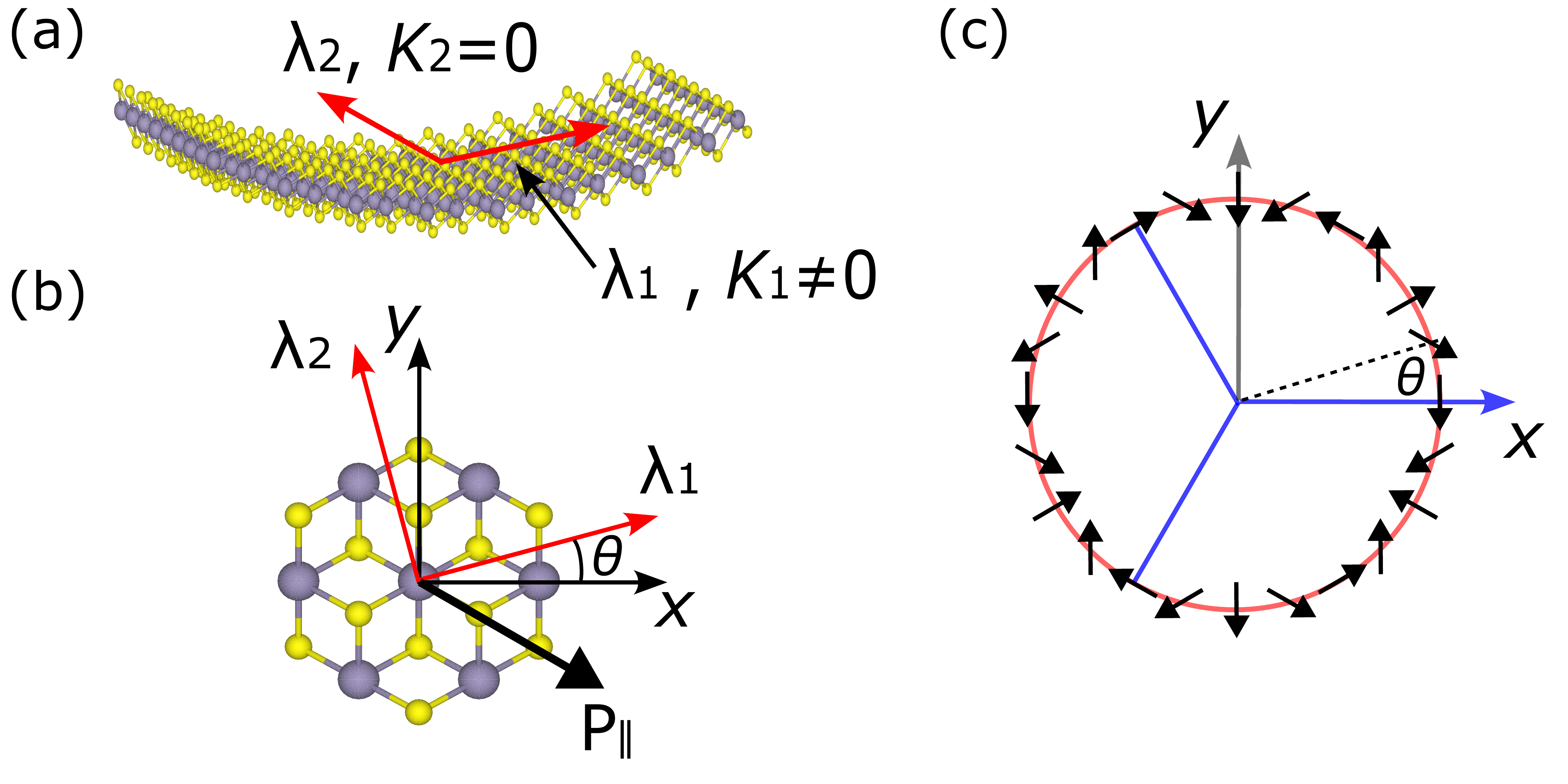}\\
  \caption{(a)Schematic illustration of a slab bent along the (principal) direction $\bm{\lambda}_{1}$ with a (principal) curvature $K_{1}$. Red arrows indicate the principal bending directions. 
  (b) 2D geometry for SnS$_{2}$.
   $\theta$ indicates the angle between the axis $x$ and the principal direction $\bm{\lambda}_{1}$.  The induced in-plane polarization $P_{\parallel}$ is indicated for an angle $\theta = 15 ^{\circ}$.(c) Orientation evolution of the induced in-plane polarization with respect to the bending direction ($\theta$). Threefold symmetry is highlighted by use of blue lines.} 
\label{Fig1}
\end{center}
\end{figure}

Our goal is to express the flexoelectric coefficients in terms of a ``strict-2D''
model of the layer, as a surface polarization~\cite{Zhou-15} response to curvature.
The geometry of the problem is conveniently specified via a vector function of two 
variables, ${\bf r} = {\bf r}(u,v)$, where ${\bf r}$ is a point in three-dimensional (3D) 
space in the deformed configuration, and $uv$ refers to a local parametrization of the
reference state~\cite{DoCarmo-16}.
Then the Cartesian components of the surface polarization can be expressed as 
${\bf P} = ( P_u {\bf r}_u + P_v {\bf r}_v) / \sqrt{g}$, where ${\bf r}_{\alpha} = \partial {\bf r}/\partial \xi^\alpha$ 
are deformation gradients, $\xi^\alpha = u,v$ are curvilinear coordinates, and 
$g$ is the determinant of the (2D) metric tensor $g_{\alpha \beta}$. 
(From now on, we shall use Greek indices for the in-plane $uv$ directions.)
The local curvature, in turn, is well described by the matrix elements 
of the \emph{second fundamental form}, $b_{\alpha \beta} =  {\bf r}_{\alpha \beta} \cdot {\bf n}$, where
the subscripts $\alpha \beta$ stand for (second) differentiation with respect to $u,v$ and 
${\bf n}(u,v)$ is the normal to the oriented surface. (We choose it in such a way
that ${\bf r}_u$, ${\bf r}_v$ and ${\bf n}$ form a right-handed set.)
This, in principle, allows us to write the flexoelectric coefficient in the
reference configuration as derivatives of $P_\alpha$ with respect to $b_{\beta \gamma}$.  

Real materials, however, are ``quasi-2D'', i.e., they are extended in plane but have
a nonvanishing thickness along the surface normal ($z$). Consequently, in practical
calculations one needs to assume a 3D--3D mapping ${\bf r}({u,v,z})$, involving a region 
of space that is thin but finite along $z$. We shall set ${\bf r}_z = {\bf n}$ henceforth;
then, the matrix elements of ${\bf b}$ coincide with the
Christoffel symbol $\Gamma_{z\beta \gamma}$ and relate to the $z$-gradient of the metric
as follows,   
\begin{equation}
b_{\beta \gamma} = \Gamma_{z \beta\gamma} = -\frac{1}{2}\frac{\partial g_{\beta\gamma}}{\partial z}.
\end{equation}
The latter two quantities, with the physical meaning of strain
gradients, are uniform along $z$ and are therefore the appropriate 
macroscopic variable to describe curvature in a quasi-2D context.~\cite{springolo2021direct}
The strict-2D limit is eventually recovered by integrating along $z$ the 
microscopic polarization response.

In the linear regime, the gradient of the metric coincides 
with the ``type-II'' representation of the strain-gradient 
tensor, $\frac{1}{2}{\partial g_{\beta\gamma}}/{\partial z}\simeq {\partial \varepsilon_{\beta\gamma}}/{\partial z} = \varepsilon_{\beta\gamma,z}$,
where $\varepsilon_{\beta\gamma}$ is the symmetrized infinitesimal strain.
Thus, we define the relevant flexoelectric coefficients as
first derivatives of the surface polarization, ${\bf P}$, with
respect to $\varepsilon_{\beta\gamma,z}$,
\begin{equation}
 P_{\alpha} = \mu^{2\rm{D}}_{\alpha z,\beta\gamma} \varepsilon_{\beta\gamma ,z},
\label{P_vs_eps}
\end{equation}
The derivative is intended to be taken at zero macroscopic electric field, 
corresponding to short-circuit (SC) boundary conditions;
this is the most natural choice given the extended nature of the crystal in 
the $xy$ plane.
Our choice of notation is motivated by the obvious analogy with the
definition of the type-II flexoelectric coefficients in 3D;~\cite{artlin}
one should keep in mind, however, that here 
$\textbf{P}$ is defined as a charge per unit length; 
consequently, the flexoelectric coefficients $\mu^{2\rm{D}}_{\alpha z,\beta\gamma}$ 
have the dimension of a charge.

\begin{figure} 
\begin{center}
 \includegraphics[width=1.0\columnwidth]{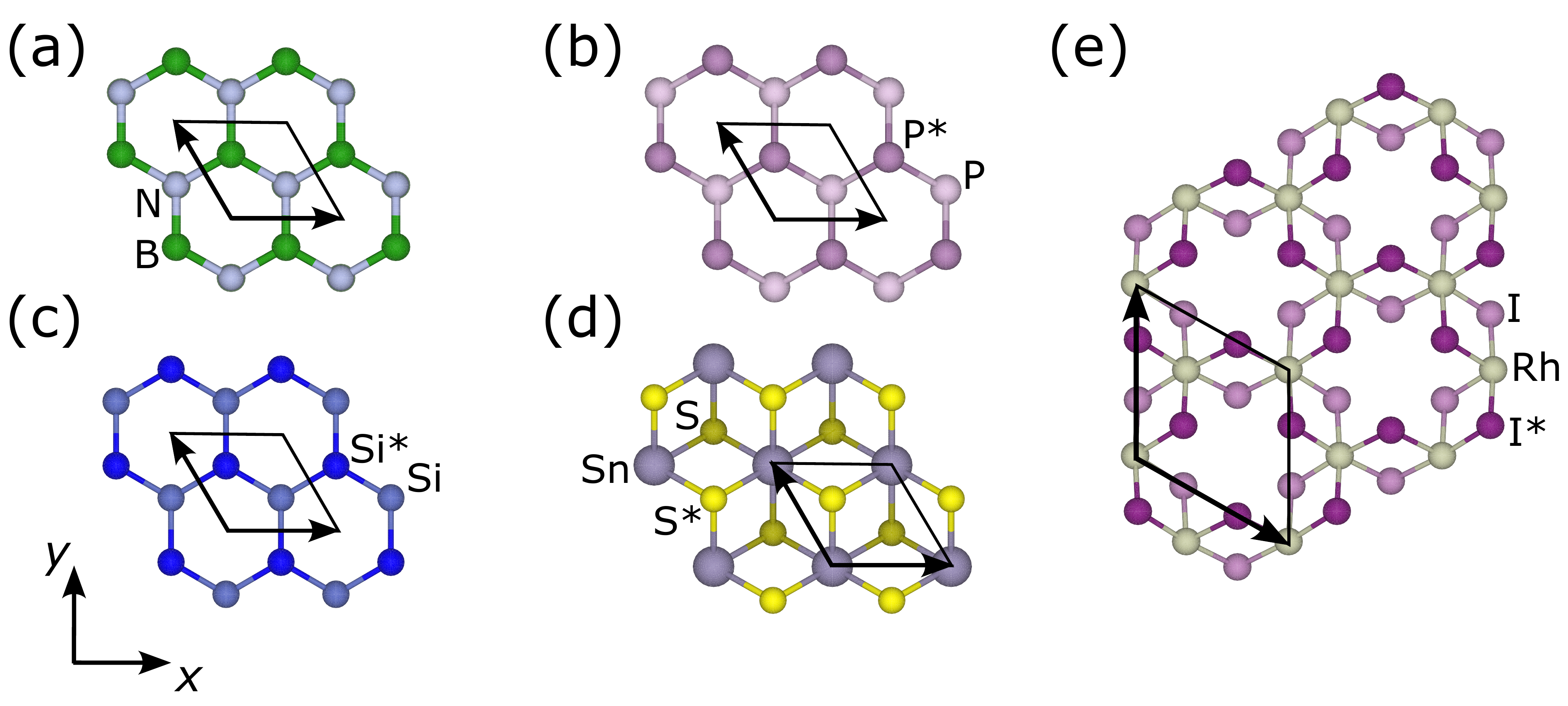}\\
  \caption{Top view of the crystal structures employed in the
linear-response calculations for the materials considered. 
The primitive cell is highlighted. 
Upper and lower atoms of the same type are indicated with the same color but different shade. Upper atoms are explicitly indicated with an asterisk.
} 
\label{Fig_2D_cells}
\end{center}
\end{figure}

A necessary condition for any of these coefficients to be nonzero is the
lack of a $z$-oriented mirror plane; similarly, $xy$ inversion cannot be a
symmetry, as the tensorial dependence on the in-plane indices is odd.
While our theory is valid in the most general case, in the following 
we focus on a subset of these materials that have the highest
symmetry compatible with the above criteria. 
More specifically, we consider crystals with a $z$-oriented three-fold axis,
space inversion symmetry and a vertical mirror plane (space groups 
P$\bar{3}m1$ and P$\bar{3}1m$,  $D_{3d}$ point group).
Within this set, 
there are only three non-zero in-plane
components of the type $\mu^{2\rm{D}}_{\alpha z,\beta\gamma}$.
Assuming a $zy$ mirror-plane ($x$ is the armchair direction in P$\bar{3}1m$ crystals,
and zigzag in P$\bar{3}m1$), they are related to each other as
\begin{equation}
\mu^{2\rm{D}}_{yz,yy} = -\mu, \qquad \mu^{2\rm{D}}_{yz,xx} = \mu^{2\rm{D}}_{xz,xy} = \mu,
\label{mu_entries}
\end{equation}
where $\mu$ is the only independent component. 
This choice leaves the freedom for two nonequivalent configurations, related by a $z \rightarrow -z$ mirror operation or, equivalently, by a rotation of $(2n+1)\pi/3$ about the threefold axis.  
The two structural variants have opposite $\mu$ coefficient, which results
in a sign ambiguity unless the atomic positions are fully specified; our
data refer to the structures shown in Fig.~\ref{Fig_2D_cells} and reported 
in the Supplemental Material~\cite{supplemental_prl}.

To see the practical implications of Eq.~(\ref{mu_entries}), it is convenient to 
represent the strain-gradient components in terms of principal curvatures ($K_i$) and directions 
($\bm{\lambda}_i$) as $\varepsilon_{\beta\gamma,z}= \sum_{i} K_{i} \lambda^{\beta}_{i} \lambda^{\gamma}_{i}$.
Then, by writing $\bm{\lambda}_1 = [\cos{(\theta)},\sin{(\theta)}]$ and $\bm{\lambda}_2 = [-\sin{(\theta}),\cos{(\theta)}]$ in terms of the angle $\theta$
[Fig.~\ref{Fig1}(b)], we find
\begin{equation}
\textbf{P}_{\parallel}(\theta) = \mu \left( \sin(3\theta) \bm{\lambda}_{1} - \cos(3\theta)\bm{\lambda}_{2} \right)(K_{1}-K_2). 
\label{lambda_P_flexo}
\end{equation}  
Remarkably, 
the modulus of the linearly induced polarization is 
$\theta$-independent, ${P}_{\parallel}(\theta) = \vert\mu (K_1-K_2)\vert$
and only depends on the difference between the principal curvatures.
(The case $K_1=K_2$ corresponds to a deformation with spherical 
symmetry, which locally preserves the threefold axis of the structure
and therefore cannot produce an in-plane ${\bf P}$.)
On the other hand, the polarization direction continuously rotates in plane
with an angular periodicity that is three times that of the principal axes.
This means that the induced polarization is manifestly invariant with respect to 
$\theta \rightarrow \theta \pm m2\pi/3$ [Fig.~\ref{Fig1}(c)] with integer $m$,   
again consistent with the $D_{3d}$ symmetry of the crystal.

Such a peculiar angular dependence of the polarization response is common to many 
electromechanical properties of hexagonal layers.
For example, an analogous behavior was found in Naumov \textit{et al.}~\cite{naumov2009unusual} 
for the polarization induced at the second-order in the curvature in a monolayer of h-BN.
Most importantly,  
Eq.~\eqref{mu_entries} matches the internal symmetries of the 
\emph{piezoelectric} tensor, $e_{\alpha,\beta\gamma}$, in BN and other isostructural materials (e.g., MoS$_{2}$) with 
a $z$-mirror plane but no space inversion.
Indeed for these materials we have  
\begin{equation}
e_{y,yy} = -E, \qquad e_{y,xx} = e_{x,xy} = E,
\end{equation} 
and the induced polarization
can be written exactly as in Eq.~(\ref{lambda_P_flexo}), provided that we replace the principal curvatures
 $K_i$ with the principal stretches $\varepsilon_i$ and the flexoelectric constant $\mu$ with $E$.
This analogy suggests that a flexoelectric effect with the characteristics of 
Eq.~(\ref{lambda_P_flexo}) should be present in a bilayer of h-BN and related compounds, provided
that they are appropriately stacked (AA' or AB') in order to recover $D_{3d}$ symmetry.
In this case, within the assumption that the layers do not interact significantly,
one expects
\begin{equation}
\mu \simeq E h,
\label{Leh}
\end{equation}
where $h$ is the interlayer distance. Eq.~(\ref{Leh}) generalizes
the results of Ref.~\onlinecite{duerloo2013flexural} to a generic flexural deformation;
here, we shall use it as a validation of our computational method by including the h-BN bilayer
in our materials test set.

We shall focus now on the first-principles calculation of $\mu^{2\rm{D}}_{\alpha z,\beta\gamma}$.
Our calculations are performed within the local-density approximation
as implemented in the ABINIT package~\cite{Gonze2009,romero-20}.
We represent the 2D crystals by using the primitive 
surface cells shown in Fig.~\ref{Fig_2D_cells}; we assume periodic boundary conditions along $z$, with 
a reasonably large separation, $L$, between images.
Following Ref.~\cite{springolo2021direct}, we express
the flexoelectric coefficients in terms of a clamped-ion (CI) 
and a lattice-mediated (LM) contributions, 
$\mu^{2\rm{D}}_{\alpha z,\beta\gamma} = \mu^{2\rm{D,CI}}_{\alpha z,\beta\gamma} + \mu^{2\rm{D},\rm{LM}}_{\alpha z,\beta\gamma}$.
We find~\cite{supplemental_prl} 
\begin{equation}
\begin{split}
{\mu}^{2\rm{D,CI}}_{\alpha z,\beta\gamma} = &- \dfrac{1}{2} L \left(\bar{\mu}^{\rm{bulk}}_{\alpha \gamma,z\beta} + \bar{\mu}^{\rm{bulk}}_{\alpha \beta,z\gamma}\right) + \\
& -\dfrac{1}{2} L \sum_{\kappa} \tau_{\kappa z}\left( \bar{P}^{(1,\beta)}_{\alpha,\kappa \gamma} + \bar{P}^{(1,\gamma)}_{\alpha,\kappa \beta} \right),
\end{split}
\label{CI_2D_mu}
\end{equation}
where $\bm{\tau}_{\kappa}$ is the undistorted location of sublattice $\kappa$,
and 
$\bar{\textbf{P}}^{(1,\beta)}_{\kappa \gamma}$ is
the first-order spatial dispersion (along $\beta$) of the 
macroscopic polarization response to a displacement of the atom $\kappa$ along the direction $\gamma$~\cite{artlin,Royo-22}.
Similarly, we write the LM contribution as
\begin{equation}
\mu^{2\rm{D},\rm{LM}}_{\alpha z,\beta\gamma} = \dfrac{1}{S} Z^{(\alpha)}_{\kappa\rho}\mathcal{L}^{\kappa}_{\rho z,\beta\gamma},
\label{LM_2D_mu}
\end{equation}
where $Z^{(\alpha)}_{\kappa\rho}$ are the Born effective charges and $\bm{\mathcal{L}}$ is the flexoelectric internal-relaxation 
tensor of the isolated layer as defined and calculated in Ref.~\cite{springolo2021direct}.

All ingredients in Eqs.~(\ref{CI_2D_mu}) and (\ref{LM_2D_mu})
can be readily calculated with the linear-response and long-wave~\cite{PhysRevX.9.021050,romero-20,Royo-22} modules of ABINIT.
Computational parameters and convergence tests are described in the Supplemental Material~\cite{supplemental_prl}, 
where we also discuss and prove the following formal results: 
(i) the contribution of the metric variation to $\mu^{2\rm{D}}_{\alpha z,\beta\gamma}$ vanishes, unlike the 
out-of-plane case studied in Ref.~\onlinecite{springolo2021direct}; (ii) the type-II 2D coefficients described here
directly relate to the macroscopic 3D coefficients of the supercell in type-I~\cite{artlin} form 
via $\mu^{2\rm{D}}_{\alpha z,\beta\gamma} = -L \mu^{\rm I}_{\alpha z,\beta\gamma}$, without
the need for calculating additional terms; (iii) modifying the electrostatic kernel to account for 
the 2D nature of the problem~\cite{Royo2021} is unnecessary as long as the macroscopic dielectric tensor is 
diagonal. (This is the case for all materials considered here.)
Note that the $\mu^{2\rm{D}}_{\alpha z,\beta\gamma}$ coefficients that we define and calculate are 
intrinsic properties of the extended 2D crystal; therefore they are applicable to modeling an arbitrary
rippled or bent structure, as we shall illustrate (and substantiate with explicit validation tests) 
shortly.

As a first consistency check of our method, we shall prove that 
we recover Eq.~(\ref{Leh}) and, therefore, the results of Ref.~\onlinecite{duerloo2013flexural} in the case of the h-BN bilayer.
To perform this test, we first calculate the value of the piezoelectric constant for an isolated h-BN monolayer,
obtaining $E^{\rm CI}=0.1230$ \textit{e}/bohr and $E^{\rm LM}=-0.0810$ \textit{e}/bohr for the CI and LM part, respectively, yielding a total $E=0.0420$ \textit{e}/bohr.
Then, we relax the geometry of an AA'-stacked bilayer, consistent with the geometry used in Ref.~\onlinecite{duerloo2013flexural}, obtaining an interlayer distance of $h= 6.14$ bohr [the experimental bulk spacing is $6.2928$ bohr \cite{X-rayBN}].
By plugging these values into Eq.~(\ref{Leh}) we obtain $\mu^{\rm CI}=0.7548$ \textit{e}, $\mu^{\rm LM}=-0.4970$ \textit{e} and $\mu=0.2577$ \textit{e},
in excellent agreement with the results reported in Table~\ref{mu_CI_LM_RI}.

The calculated 2D in-plane flexocoefficients ($\mu^{2\rm{D}}_{\alpha z,\beta\gamma}$) of silicene, blue phosphorene, SnS$_2$ and RhI$_3$ monolayers, as well as of the h-BN bilayer with AA' stacking, are shown in Table \ref{mu_CI_LM_RI} along with the corresponding out-of-plane ones.
[The latter calculated as $\mu^{2\rm{D}}_{zz,\beta\beta} = \varphi/\epsilon_{0}$, from the flexovoltage $\varphi$ as defined in Ref.~\onlinecite{springolo2021direct}]
Interestingly, our results indicate that the in-plane flexoelectric response is much larger than the out-of-plane one, in several cases about one order of magnitude larger.
This trend is common to all the studied materials in spite of the systematic sign reversal observed for the CI and LM contributions; 
both are generally comparable in module, except for RhI$_{3}$ where the LM term is one order of magnitude smaller.

\begin{table} 
\setlength{\tabcolsep}{6pt}
\begin{center}

\begin{tabular}{c|r|r|r|r}\hline\hline
    \T\B & \multicolumn{1}{c}{} & \multicolumn{1}{c}{$\mu^{2\rm{D}}_{yz,xx}$} & \multicolumn{1}{c}{} & \multicolumn{1}{|c}{$\mu^{2\rm{D}}_{zz,xx}$} \\ \hline
    \T\B & \multicolumn{1}{c}{CI} & \multicolumn{1}{c}{LM} & \multicolumn{1}{c}{RI} & \multicolumn{1}{|c}{RI} \\ \hline
   Si \T\B  & 0.0299  & 0.0000 & 0.0299 & 0.0032 \\
	BN(bilayer) \T\B  & 0.7569  & $-$0.4947  & 0.2622 & $-$0.0304  \\
	(blue)P \T\B   & 0.0721  & 0.0000  & 0.0721 & 0.0212 \\ 
	SnS$_{2}$ \T\B & 0.1257  & $-$0.2263  & $-$0.1006 & 0.0198 \\
	RhI$_{3}$ \T\B & $-$0.1766 & 0.0102 & $-$0.1664 & $-$0.0062 \\
  \hline \hline
\end{tabular}
\caption{Characteristic 2D flexocoefficients due to a flexural deformation. Three left-most columns show the Clamped-Ion(CI), Lattice-Mediated(LM), and Relaxed-Ion(RI) contributions to the in-plane response $\mu^{2\rm{D}}_{yz,xx}$. Fourth column shows the out-of-plane RI response, $\mu^{2\rm{D}}_{zz,xx}$. Results are provided in units of electronic charge.}
\label{mu_CI_LM_RI}
\end{center}
\end{table}

To get a flavor of how large the predicted effect is, we can compare it to the nonlinear one described by Naumov {\em et al.}~\cite{naumov2009unusual}, which involves the in-plane polarization response to the square of the local curvature.
Of course, the coefficients are not directly comparable as they occur at different orders in the deformation amplitude; however, we can estimate a critical radius of curvature ($R^*$) at which the respective magnitudes of the induced polarization become similar.
Extracting a second-order coefficient from the numbers provided in Ref.~\onlinecite{naumov2009unusual} and employing the linear-order coefficient for the h-BN bilayer, we obtain an unrealistically small 
value of $R^{*} \sim 0.75$ bohr, with the linear effect prevailing at any $R>R^{*}$.
For instance, Naumov {\em et al.} obtain induced polarizations  
$P \sim 0.013$ e/bohr for an h-BN monolayer corrugated along a zigzag direction with an average radius  $R \sim 6.7$ bohr and at a CI level.
Instead, taking our CI coefficient for the h-BN bilayer [Table~\ref{mu_CI_LM_RI}], the present linear effect 
reaches the same magnitude at curvature radii $R \sim 60$ bohr.
At deformation regimes that are currently attainable in a laboratory~\cite{wang-19} ($R \sim 10$ $\mu$m), 
the linear response of the bilayer is stronger than the nonlinear one of the monolayer 
by five orders of magnitude.

\begin{figure}
\centering
\includegraphics[height=1.4in]{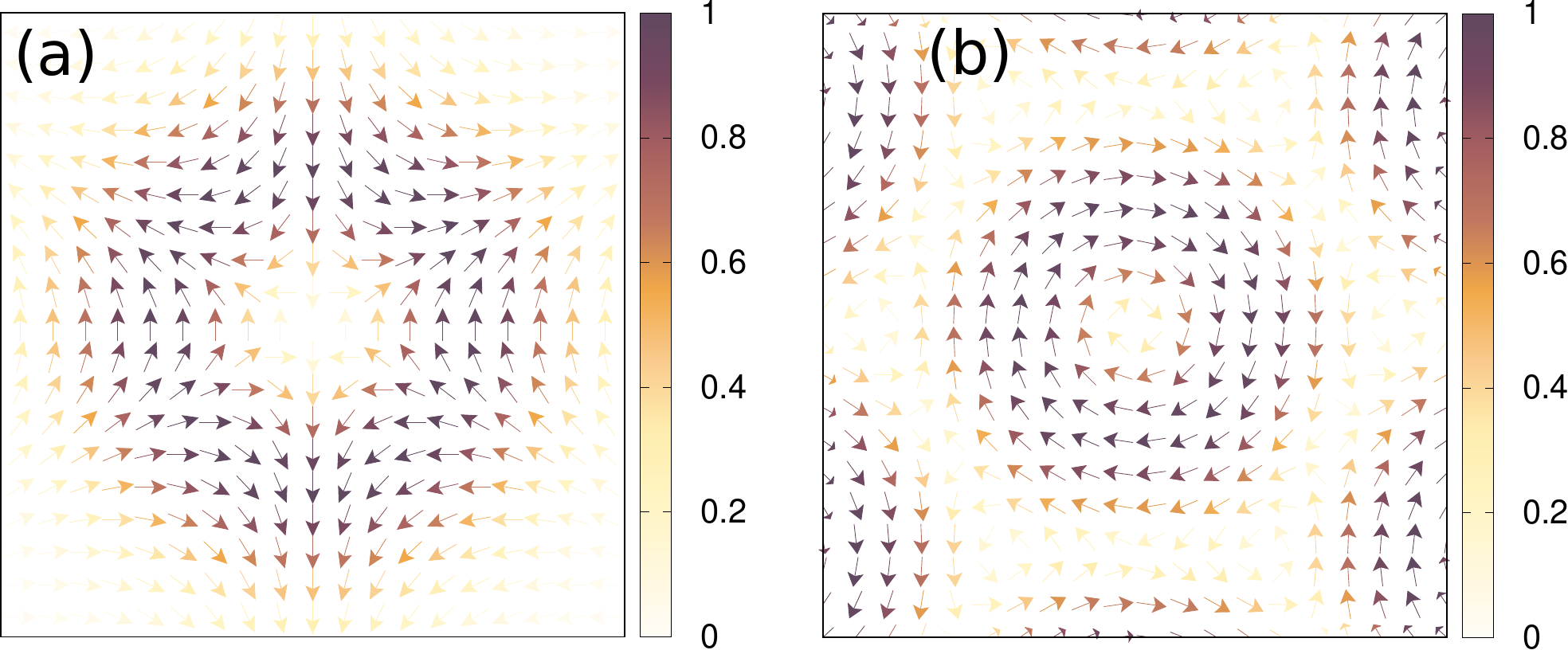}
\caption{ Polarization textures associated with two different ripple patterns. (a): Gaussian bump of the type 
$z=Ae^{-(x^2+y^2)/{\sigma}^2}$. (b): Periodic pattern of the type $z=A\sum_{i=1}^3 \sin({\bf q_i}\cdot {\bf r})$, with
${\bf q}_1 = q(1,0,0)$, ${\bf q}_2 = q(-1/2,\sqrt{3}/2,0)$, ${\bf q}_3 = q(-1/2,-\sqrt{3}/2,0)$ and $q=2\pi/L$. The arrows indicate the polarization 
direction, its amplitude ({in units of $\vert P^{\rm{max}}\vert =1.48A\mu/\sigma^2$ and $\vert P^{\rm{max}}\vert=1.75A\mu q^2$ respectively)} is defined by the color scale. }
\label{fig3}
\end{figure}

A direct outcome of the predicted flexoelectric coupling is the emergence of topologically
nontrivial in-plane polarization textures in arbitrary rippled states.
%
To illustrate this, in Fig.\ref{fig3}(a) we show the in-plane polarization field that results from
an isolated Gaussian-shaped bump, 
consisting in a textbook antivortex structure with a topological charge (winding number)~\cite{JunqueraRMP23} of ${Q}=-2$.
In Fig.\ref{fig3}(b) we consider a periodic deformation pattern, consisting
of three superimposed sinusoidal ripples with threefold symmetry;
the result is a hexagonal lattice of clockwise vortices (${Q}=1$).
By playing with these two examples~\cite{supplemental_prl} we could obtain a whole range of patterns, including
an array of monopoles, and in-plane textures that closely resemble those of Ref.~\cite{bennett2023polar}.
All these structures could, in principle, be characterized experimentally by optical means 
(e.g., via second harmonic generation~\cite{Zhang_2020}).

As a corollary of the above, our theory predicts an axial spontaneous polarization in nanotubes made of 
2D trigonal crystals. 
In particular, a nanotube 
constructed by folding a layer along $\bm{\lambda}_1$ acquires a linear 
polarization in the form
\begin{equation}
P^{\rm 1D} = -2\pi \mu \cos(3\theta),
\label{P_tube}
\end{equation}
independent of its radius, $R$.
%
%
To test this result, we perform explicit ground-state DFT calculations of zigzag SnS$_2$ nanotubes via 
VASP.5.4.~\cite{Kresse-93,Kresse-94,Kresse-96,Kresse-99}. We find~\cite{supplemental_prl} excellent agreement between
the computed Berry phase polarization of the tube and the predictions of Eq.(\ref{P_tube}). 
Note that a circulating (azimuthal) component of the polarization, 
of amplitude $P_\theta = \mu /R \sin(3\theta)$, is generally also present. 
$P_\theta$ is largest for $\theta=(2n+1)\pi/6$, with integer $n$, where
Eq.(\ref{P_tube}) yields a vanishing result. For $\theta \neq n\pi/6$
the two components coexist, and the polarization field becomes chiral.

Another central consequence of our theory is that the present mechanism endows flexural phonons with longitudinal
electric fields, which may be of considerable importance in the context of electron-phonon couplings.
For a propagation direction ${\bf q} = q \bm{\lambda}_1$, a combination of Eqs.~\eqref{CI_2D_mu} and~\eqref{LM_2D_mu}
with the formalism developed in Refs.~\onlinecite{Royo2021,ponce2022drift,ponce2022long} leads to a macroscopic electrostatic
contribution to the scattering potential (and hence to the diagonal electron-phonon matrix elements) 
in the form
\begin{equation}
V^\parallel({\bf q}) = -2\pi i q^2 \mu \sin(3\theta) + O(q^3).
\label{eq_scatpot}
\end{equation}
An independent proof of this formula,
based on the formalism of Ref.~\cite{Royo2021}, is provided in the Supplemental Material~\cite{supplemental_prl}.
%
%
%
As the ensemble average of the flexural phonon amplitude diverges at any temperature ~\cite{Fischetti2016,Gunst2017}
for ${\bf q}\rightarrow 0$, we suspect that Eq.~(\ref{eq_scatpot}) may contribute significantly 
to electron transport; a verification of this point will be an interesting topic for
future studies.

To summarize, we have predicted and numerically demonstrated the existence of an in-plane polarization response to a flexural deformation 
in trigonal 2D monolayers. 
We have also discussed the topological implications of our results in a variety of rippled and bent geometries.
%
%
%
In addition, we have demonstrated the relevance of the effect in dealing with the long-range electrostatic fields induced by a flexural phonon, 
with potential implications for electron-phonon physics.
We hope that our results will stimulate experimental efforts at detecting the effects we describe here, e.g., along
the lines that we suggest in Section 8 of Ref.~\onlinecite{supplemental_prl}.
\\
\begin{acknowledgments}
 We acknowledge support from Ministerio de Ciencia e Innovaci\'on (MICINN-Spain) through Grant No. PID2019-108573GB-C22; from Severo Ochoa FUNFUTURE center of excellence (CEX2019-000917-S); from Generalitat de Catalunya (Grant No. 2021 SGR 01519); and from the European Research Council (ERC) under the European Union's Horizon 2020 research and innovation program (Grant Agreement No. 724529).
 Part of the calculations were performed at the Supercomputing Center of Galicia (CESGA).
\end{acknowledgments}

\bibliography{refs1}

\end{document}


\maketitle

\tableofcontents

\section{Introduction}

Here we provide additional material in support of our results.
In Sec.~\ref{def_flexoresponse}, we prove
the absence of a metric contribution 
to the in-plane coefficients considered here, in contrast to the out-of-plane case 
studied in our previous work~\cite{springolo2021direct}.
In Sec.~3, we provide a detailed derivation of the in-plane 2D flexoelectric coefficients in their type-II form. 
Then, we split the total (relaxed-ion) response into clamped-ion and lattice-mediated contributions, and show its relation with the 3D type-I flexoelectric tensor.
In Sec.~4, we provide the detailed computational parameters used in the primitive cell linear-response, while in Sec.~5 we describe the computational method employed for the calculation of the axial polarization within SnS$_{2}$ nanotubes, and discuss the related outcomes.
In Sec.~6, we show that the in-plane flexoelectric coefficients 
also provide an exact description of the longitudinal macroscopic electric fields associated to a flexural phonon. 
In Sec.~7 the impact of the choice of the exchange-correlation functional on the quantity considered in this work is discussed.
Sec.~8 is assigned to the illustration of two experimental setups that could possibly confirm the theoretical predictions and results presented in the main text.
Finally, in Sec.~9, the non trivial polarization textures associated with several types of ripples are presented.

\section{Flexoelectric response of 2D materials\label{def_flexoresponse}}

The flexoelectric response of a quasi-2D material can be either defined as the derivative (with respect to the uniform strain-gradient $\varepsilon_{\beta\gamma,\lambda}$ describing the deformation considered) of the electrostatic potential $V$ at fixed macroscopic displacement field $\textbf{D}$ (open circuit) or as the derivative of the macroscopic $\textbf{D}$ at fixed $V$ 
(short circuit). 
While the former definition is appropriate for the out-of-plane response~\cite{springolo2021direct}, the latter is the natural choice for the in-plane components studied here, given the extended nature of the crystal therein.
Following the prescriptions of Refs.~\cite{stengel-13b,chapter-15}, we define the flexoelectric response via a curvilinear-coordinate representation of the microscopic electric displacement field via 
\begin{equation}
\hat{\textbf{D}}^{(1)}(\textbf{r}) = \hat{\bm{\epsilon}}^{(0)} \cdot \hat{\textbf{E}}^{(1)}(\textbf{r})  + \hat{\bm{\epsilon}}^{(1)}(\textbf{r})\cdot\hat{\textbf{E}}^{(0)}(\textbf{r}) + \hat{\textbf{P}}^{(1)}(\textbf{r}).
\label{curv_D}
\end{equation}
(The hat symbol indicates the curvilinear representation of the relevant quantities and the $(1)$ superscript indicates first order in the strain gradient.)
The second term in the rhs of Eq.~(\ref{curv_D}), $\hat{\bm{\epsilon}}^{(1)}(\textbf{r})\cdot\hat{\textbf{E}}^{(0)}(\textbf{r})$, 
orginates from the \textit{geometric} variation of the vacuum permittivity 
$\hat{\bm{\epsilon}}(\textbf{r}) = \epsilon_{0}\sqrt{g}\hat{\textbf{g}}^{-1}(\textbf{r})$ ~\cite{chapter-15}, and couples 
the linear variation of the curvilinear microscopic electric displacement with the microscopic electric field within the undistorted configuration, $\hat{\textbf{E}}^{(0)}(\textbf{r}) = \textbf{E}^{(0)}(\textbf{r})$ ($\textbf{E}^{(0)}(\textbf{r})$ being the Cartesian representation of the electric field) through the linear variation of the metric tensor $\hat{\textbf{g}}(\textbf{r})$ and its determinant $g$.
This means that, in principle, a metric contribution may be present (and hence the
first-order $\hat{\bf D}$ may differ from the first-order $\hat{\bf P}$) even if 
short-circuit  electrical boundary conditions are assumed.
However, one can show that, for the in-plane response considered in this work, the geometric contribution vanishes. 
Indeed the linear variation of the metric contribution, $\hat{\bf D}^{(\rm{met})}= \hat{\bm{\epsilon}}^{(1)}(\textbf{r})\cdot\hat{\textbf{E}}^{(0)}(\textbf{r})$, due to a uniform strain-gradient of the form $\varepsilon_{\beta\gamma,z}$ takes the form ~\cite{chapter-15}
\begin{equation}
\dfrac{\partial \hat{D}^{(\rm{met})}_{\alpha} (\textbf{r})}{\partial \varepsilon_{\beta\gamma,z}}=  \epsilon_{0} z \left( \delta_{\beta\gamma}E^{(0)}_{\alpha}(\textbf{r}) - \delta_{\alpha\beta}E^{(0)}_{\gamma}(\textbf{r}) - \delta_{\alpha\gamma}E^{(0)}_{\beta}(\textbf{r})  \right) .
\end{equation}
and its in-plane components ($\alpha = x$, or $y$) are zero when integrated on the volume of the unit-cell, leading to 
\begin{equation}
\langle\hat{\bf D}^{\rm{SC}}_{\parallel}\rangle = \langle\hat{\bf D}_{\parallel}\rangle\vert_{\langle\hat{\bf E}^{(1)}_{\parallel}\rangle=0} = \langle\hat{\bf P}_{\parallel}^{(1)}\rangle\vert_{\langle\hat{\bf E}^{(1)}_{\parallel}\rangle=0}
\end{equation}
once the unit-cell average of the in-plane response is considered and SC conditions ($\langle\hat{\bf E}^{(1)}_{\parallel}\rangle = 0$) are applied.

\section{In-plane response to a flexural deformation}

The microscopic \textit{curvilinear} polarization linear-response along an in-plane (curvilinear) direction ($\alpha = x$ or $y$) due to a symmetrized uniform strain-gradient of the type $\varepsilon_{\beta\gamma,\lambda}$ can be written as~\cite{stengel-13b,chapter-15} 
\begin{equation}
\hat{P}^{(1)}_{\alpha}(\textbf{r}) = P^{(\rm{U})}_{\alpha,\beta\gamma}(\textbf{r})\varepsilon_{\beta\gamma}(\textbf{r}) + P^{(\rm{G})}_{\alpha\lambda,\beta\gamma}(\textbf{r})\varepsilon_{\beta\gamma,\lambda}
\end{equation}
where summation over repeated indices is implied, and  
\begin{subequations}
\begin{align}
 P^{(\rm{U})}_{\alpha,\beta\gamma}(\textbf{r}) = & \underbrace{- \dfrac{1}{2}\sum_{k} \left( P^{(1,\gamma)}_{\alpha,\kappa\beta}(\textbf{r}) + P^{(1,\beta)}_{\alpha,\kappa\gamma}(\textbf{r})\right)}_{\rm{Clamped-Ion}} + \underbrace{P^{(0)}_{\alpha ,\kappa\rho}(\textbf{r})\Gamma^{\kappa}_{\rho ,\beta\gamma}}_{\rm{Lattice-Mediated}},  \label{eq_PsupU} \\
 P^{(\rm{G})}_{\alpha\lambda,\beta\gamma}(\textbf{r}) = & \underbrace{\dfrac{1}{2} \sum_{\kappa} \left( P^{(2,\gamma\lambda)}_{\alpha ,\kappa\beta}(\textbf{r}) + P^{(2,\lambda\beta)}_{\alpha ,\kappa\gamma}(\textbf{r}) - P^{(2,\beta\gamma)}_{\alpha ,\kappa\lambda}(\textbf{r}) \right)}_{\rm{Clamped-Ion}} \nonumber  \\
& - \underbrace{P^{(1,\lambda)}_{\alpha ,\kappa\rho}(\textbf{r})\Gamma^{\kappa}_{\rho,\beta\gamma}}_{\rm{Mixed}} + \underbrace{P^{(0)}_{\alpha ,\kappa\rho}(\textbf{r})L^{\kappa}_{\rho\lambda,\beta\gamma}}_{\rm{Lattice-Mediated}}. \label{eq_PsupG} 
\end{align}
\end{subequations}
Here, $\textbf{P}^{(\rm{U})}_{\beta\gamma}(\textbf{r})$ is the microscopic polarization linear-response due to a uniform symmetrized strain $\varepsilon_{\beta\gamma}$, and $\textbf{P}^{(\rm{U})}_{\beta\gamma}(\textbf{r})r_{\lambda} = \textbf{P}^{(\rm{U})}_{\beta\gamma}(\textbf{r})\dfrac{\partial \varepsilon_{\beta\gamma}(\textbf{r})}{\partial \varepsilon_{\beta\gamma,\lambda}}$ and $\textbf{P}^{(\rm{G})}_{\lambda,\beta\gamma}(\textbf{r})$ have the physical interpretation of a local piezoelectric (U) and a flexoelectric (G) (type-II) coefficient respectively due to a uniform strain-gradient of the type $\varepsilon_{\beta\gamma,\lambda} = \partial \varepsilon_{\beta\gamma}(\textbf{r})/\partial r_{\lambda}$. 
A flexural deformation can be described by a symmetrized transverse strain gradient of the type $\varepsilon_{\beta\gamma,z}$ and then the longitudinal microscopic polarization response due to a flexural deformation is
\begin{equation}
\dfrac{\partial \hat{P}_{\alpha}(\textbf{r})}{\partial \varepsilon_{\beta\gamma,z}} =  P^{(\rm{U})}_{\alpha,\beta\gamma}(\textbf{r}) z + P^{(\rm{G})}_{\alpha z,\beta\gamma}(\textbf{r}).
\end{equation}

The total (relaxed-ion) 2D longitudinal flexoelectric coefficient is finally obtained as 
\begin{equation}
\mu^{2\rm{D}}_{\alpha z,\beta\gamma} = \dfrac{1}{S} \int_{\Omega} d^{3}r \dfrac{\partial \hat{P}_{\alpha}(\textbf{r})}{\partial \varepsilon_{\beta\gamma,z}} = \dfrac{1}{S} \int_{\Omega} d^{3}r P^{(\rm{U})}_{\alpha,\beta\gamma}(\textbf{r}) z + L\mu^{\rm{II}}_{\alpha z,\beta\gamma}
\label{2D_mu_NI}
\end{equation}
where $S$ is the unit-cell surface, $L$ the out-of-plane dimension of the supercell where the layer is accomodated in, and the relation $\mu^{\rm{II}}_{\alpha z,\beta\gamma} = \dfrac{1}{\Omega} \int_{\Omega} d^{3}r P^{(\rm{G})}_{\alpha z,\beta\gamma}(\textbf{r})$~\cite{artlin,chapter-15} has been used. 
$\mu^{\rm{II}}_{\alpha z,\beta\gamma}$ is the type-II bulk flexoelectric coefficient of the supercell, the volume of the latter indicated as $\Omega$.

In the following sections we shall recast Eq.~(\ref{2D_mu_NI}) in a form that is suitable for direct implementation and that provides an exact separation between clamped-ion (CI) and lattice-mediated (LM) contributions.

\subsection{Clamped-ion 2D flexocoefficient}

The clamped-ion contribution to $\mu^{2\rm{D}}_{\alpha z,\beta\gamma}$ is formally written as
\begin{equation}
\mu^{2\rm{D},\rm{CI}}_{\alpha z,\beta\gamma} =  -\dfrac{1}{2S} \sum_{\kappa}\int_{\Omega} d^{3}r \left( P^{(1,\gamma)}_{\alpha,\kappa \beta}(\textbf{r}) z + P^{(1,\beta)}_{\alpha,\kappa \gamma}(\textbf{r}) z\right) + L \mu^{\rm{II},\rm{CI}}_{\alpha z,\beta\gamma}.
\label{2D_CI_mu_NI}
\end{equation}
Using the fact that the system is finite along the out-of-plane $z$-direction, we can write
\begin{equation}
\begin{split}
 -\dfrac{1}{S}\sum_{\kappa} \int_{\Omega} d^{3}r z P^{(1,\gamma)}_{\alpha,\kappa \beta}(\textbf{r}) & =  -\dfrac{1}{S} \sum_{\kappa} \int_{\Omega} d^{3}r \left(z-\tau_{\kappa z} \right) P^{(1,\gamma)}_{\alpha,\kappa \beta}(\textbf{r}) - \dfrac{1}{S}\sum_{\kappa} \tau_{\kappa z}\int_{\Omega} d^{3}r P^{(1,\gamma)}_{\alpha,\kappa \beta}(\textbf{r}) \\
 & = - \dfrac{1}{S}\sum_{\kappa}\int_{\Omega} d^{3}r P^{(2,\gamma z)}_{\alpha,\kappa \beta}(\textbf{r}) - \dfrac{1}{S} \sum_{\kappa} \tau_{\kappa z} \int_{\Omega} d^{3}rP^{(1,\gamma)}_{\alpha,\kappa \beta}(\textbf{r}) \\
 & = -2 L \mu^{\rm{I},\rm{CI}}_{\alpha\beta ,\gamma z} - L \sum_{\kappa} \tau_{\kappa z} \bar{P}^{(1,\gamma)}_{\alpha,\kappa \beta}. 
\end{split}
\end{equation}
Here, the overbar indicates avarage over the supercell volume and we have used the relationship~\cite{artlin} between the type-I CI flexocoefficients and the sublattice summation of $\bar{P}^{(2,\gamma z)}_{\alpha,\kappa \beta}$.
Finally, using the relationship $\mu^{\rm{I}}_{\alpha \beta,\gamma z} = (1/2)\left( \mu^{\rm{II}}_{\alpha z,\beta\gamma} + \mu^{\rm{II}}_{\alpha \gamma,\beta z} \right)$~\cite{chapter-15,artlin} and gathering all the terms together, Eq.(\ref{2D_CI_mu_NI}) can be rewritten as
\begin{equation}
\mu^{2\rm{D},\rm{CI}}_{\alpha z,\beta\gamma} = -\dfrac{L}{2} \left[ \left(\mu^{\rm{II},\rm{CI}}_{\alpha\gamma,z\beta} + \mu^{\rm{II},\rm{CI}}_{\alpha\beta,z\gamma}\right) + \left(\bar{P}^{(1,\beta)}_{\alpha,\kappa\gamma} + \bar{P}^{(1,\gamma)}_{\alpha,\kappa\beta}\right)\tau_{\kappa z} \right].
\label{generic_2D_mu}
\end{equation}

\subsection{Lattice-mediated 2D flexocoefficient}

The remaining, no clamped-ion type, contributions to the 2D flexocoefficient of Eq.~(\ref{2D_mu_NI}) are 
\begin{equation}
\dfrac{1}{S} \int_{\Omega} d^{3}r z P^{(0)}_{\alpha ,\kappa\rho}(\textbf{r})\Gamma^{\kappa}_{\rho ,\beta\gamma} - L \bar{P}^{(1,z)}_{\alpha ,\kappa\rho} \Gamma^{\kappa}_{\rho ,\beta\gamma} + \dfrac{1}{S} Z^{(\alpha)}_{\kappa\rho} L^{\kappa}_{\rho z,\beta\gamma}.
\label{2D_mu_rel}
\end{equation}
The first term, coming from Eq.~\eqref{eq_PsupU}, can be manipulated likewise its CI counterpart, as follows
\begin{equation}
\begin{split}
\dfrac{1}{S} \int_{\Omega} d^{3}r z P^{(0)}_{\alpha ,\kappa\rho}(\textbf{r})\Gamma^{\kappa}_{\rho ,\beta\gamma} & =  \dfrac{1}{S} \int_{\Omega} d^{3}r\left( z -\tau_{\kappa z} \right) P^{(0)}_{\alpha ,\kappa\rho}(\textbf{r})\Gamma^{\kappa}_{\rho ,\beta\gamma} + \dfrac{1}{S} \tau_{\kappa z}\int_{\Omega} d^{3}r P^{(0)}_{\alpha ,\kappa\rho}(\textbf{r})\Gamma^{\kappa}_{\rho ,\beta\gamma} \\
& = \dfrac{1}{S} \int_{\Omega} d^{3}r P^{(1,z)}_{\alpha ,\kappa\rho}(\textbf{r})\Gamma^{\kappa}_{\rho ,\beta\gamma} + \tau_{\kappa z} \dfrac{1}{S} \int_{\Omega} d^{3}r P^{(0)}_{\alpha ,\kappa\rho}(\textbf{r})\Gamma^{\kappa}_{\rho ,\beta\gamma} \\
& = L \bar{P}^{(1,z)}_{\alpha ,\kappa\rho} \Gamma^{\kappa}_{\rho ,\beta\gamma} + \dfrac{1}{S} Z^{(\alpha)}_{\kappa \rho} \tau_{\kappa z} \Gamma^{\kappa}_{\rho ,\beta\gamma}.
\end{split}
\end{equation}
Now, after observing that the first term at the rhs exactly cancels the mixed one [second term in Eq.~\eqref{2D_mu_rel}], we are left with a purely lattice-mediated contribution written as
\begin{equation}
\mu^{2\rm{D, LM}}_{\alpha z,\beta\gamma} = \dfrac{1}{S} Z^{(\alpha)}_{\kappa \rho} \mathcal{L}^{\kappa}_{\rho z,\beta\gamma}
\end{equation}
where $\mathcal{L}^{\kappa}_{\rho z,\beta\gamma}$ is the internal relaxation of the $\kappa$-atom in the isolated slab due to the flexural deformation.
 The last quantity can be likewise split as
\begin{equation}
\mathcal{L}^{\kappa}_{\rho z,\beta\gamma} = \Gamma^{\kappa}_{\rho ,\beta\gamma} \tau_{\kappa z} + L^{\kappa}_{\rho z,\beta\gamma},
\end{equation}
i.e., in terms of the piezoelectric ($\bm{\Gamma}$) and flexoelectric ($\bm{L}$) internal tensors of the 3D supercell.

\subsection{Relation to the type-I bulk coefficients \label{mu2D_muI}}

In the main text we claim that the ``2D'' type-II flexoelectric coefficients defined
and calculated in this work directly relate to the macroscopic ``3D'' type-I 
flexoelectric coefficients of the supercell.
To support this statement, here we shall demonstrate the following relationship
\begin{equation}
\mu^{2\rm{D}}_{\alpha z,\beta\gamma} = -L  \mu^{\rm{I}}_{\alpha z, \beta\gamma}.
\label{eq:mu_typeI}
\end{equation}

Regarding the clamped-ion part, notice that, exploiting the relation $\Gamma^{\kappa}_{\alpha,z \beta} = \Gamma^{\kappa}_{\alpha,\beta z} = -\delta_{\alpha\beta}\tau_{\kappa z}$ valid for an isolated slab, Eq.(\ref{generic_2D_mu}) can be recast in the form
\begin{equation}
\mu^{2\rm{D},\rm{CI}}_{\alpha z,\beta\gamma} = -\dfrac{1}{2} L \left( \mu^{\rm{II},el}_{\alpha\gamma,z\beta} + \mu^{\rm{II},el}_{\alpha\beta,z\gamma} \right),
\label{eq_sum_rule}
\end{equation}
with $\mu^{\rm{II},el}_{\alpha\gamma,z\beta}$ being the type-II bulk \textit{electronic} flexoelectric tensor defined as in Ref.~\cite{Royo-22}:
\begin{equation}
\mu^{\rm{II},el}_{\alpha\gamma,z\beta} = \mu^{\rm{II,CI}}_{\alpha\gamma,z\beta} -\bar{P}^{(1,\gamma)}_{\alpha,\kappa\rho}\Gamma^{\kappa}_{\rho,z\beta}
\end{equation}

Regarding the lattice-mediated contribution, notice that in a free-standing layer the shear components of the type-I
internal-strain tensor ${\bf N}$~\cite{artlin} and the transverse components of the type-II tensor are related by  
\begin{equation}
N^{\kappa}_{\rho z,\beta\gamma} = -\Gamma^{\kappa}_{\rho ,\beta\gamma} \tau_{\kappa z} - L^{\kappa}_{\rho z,\beta\gamma} = 
-\mathcal{L}^{\kappa}_{\rho z,\beta\gamma}.
\end{equation}

Now, recalling the relationship between the type-I and type-II  representations of the flexoelectric tensor~\cite{chapter-15,artlin},
\begin{equation}
\dfrac{1}{2} \left(\mu^{\rm{II}}_{\alpha\gamma,z\beta} + \mu^{\rm{II}}_{\alpha\beta,z\gamma}\right) = \mu^{\rm{I}}_{\alpha z, \beta\gamma}
\end{equation}
we can conclude that
\begin{equation}
\mu^{2\rm{D},\rm{CI}}_{\alpha z,\beta\gamma} = -L  \mu^{\rm{I,el}}_{\alpha z, \beta\gamma}, \qquad
\mu^{2\rm{D,LM}}_{\alpha z,\beta\gamma} = -L  \mu^{\rm{I,LM}}_{\alpha z, \beta\gamma},
\end{equation}
which proves Eq.~\eqref{eq:mu_typeI}.

This result is consistent with the known fact~\cite{springolo2021direct,stengel-13b} that a 2D flexural phonon propagating in a free-standing slab is characterized by the same atomic displacement pattern as a transverse strain gradient.

\section{Computational parameters}

\begin{figure} 
\begin{center}
  \includegraphics[width=0.9\textwidth]{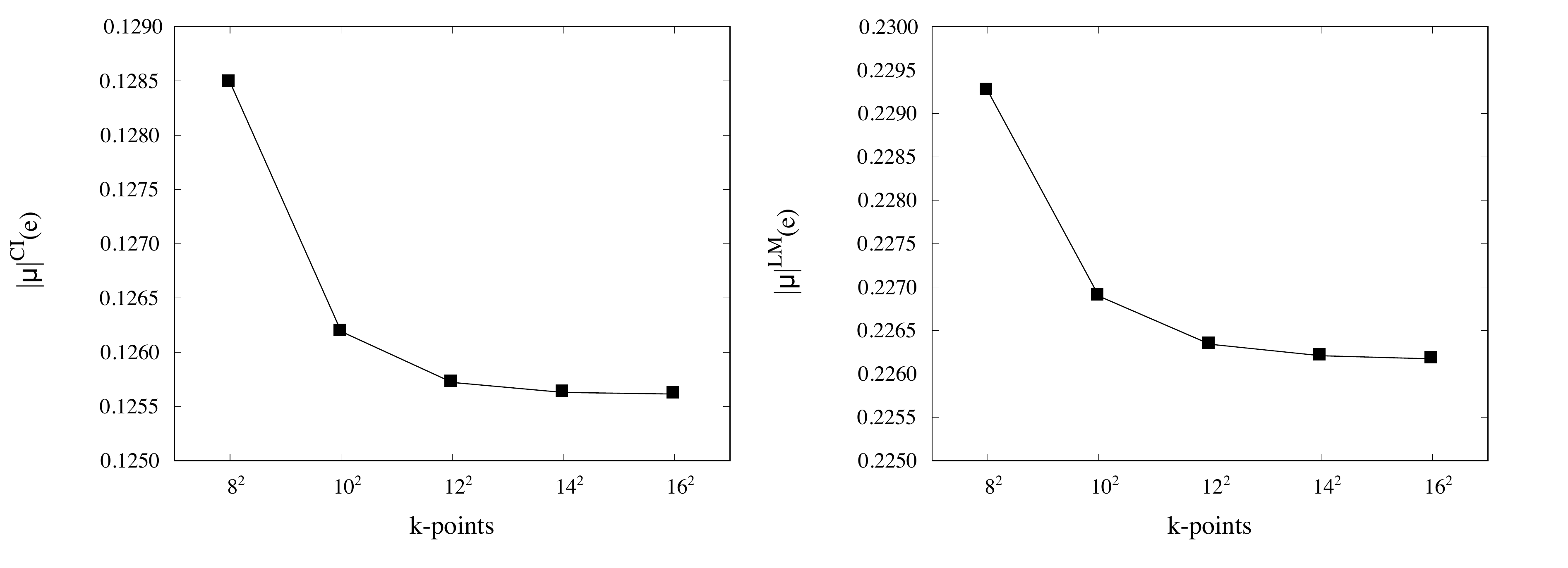}\\
  \caption{Convergence of the CI (a) and LM (b) independent component $\vert \mu \vert $ discussed in the main text as a function of the in-plane \textbf{k}-points mesh resolution, for SnS$_{2}$.} 
\label{Fig_conv}
\end{center}
\end{figure}

Norm-conserving pseudopotentials are generated with Hamann’s approach~\cite{hamann-13}, by using the “stringent” parameters of PseudoDojo~\cite{pseudodojo}, but neglecting non-linear core corrections.
We set a supercell size of $L$ = 30 bohr (15.875 \r{A}) and a plane-wave cutoff of 80 Ha; the Brillouin zone is sampled by a $\Gamma$-centered $12\times12\times2$ mesh except for silicene (a grid of $13\times13\times2$ points is used); with respect to these parameters, the calculated flexocoefficients are converged within a tolerance of 0.1 \% or better (see e.g. Fig.~\ref{Fig_conv}).
Before performing the linear-response calculations, we optimize the atomic positions and cell parameters of the unperturbed systems to a stringent tolerance ($10^{-7}$ and $10^{-5}$ atomic units for residual stress and forces, respectively); the resulting structures (detailed in Tab. \ref{Tab_struc}) are in excellent agreement with existing literature data (see e.g. Ref.~\cite{novoselov-16} and references therein).
In particular, for all the materials considered we used an hexagonal unit-cell described by primitve vectors of the form
\begin{equation}
\begin{split}
& \textbf{a}_{1} = a \begin{array}{ccc}
(1.0 &, 0.0 &, 0.0)
\end{array} \\
& \textbf{a}_{2} = b \begin{array}{ccc}
(-0.5 &, \sqrt{3}/2 &, 0.0)
\end{array} \\
& \textbf{a}_{3} = c \begin{array}{ccc}
(0.0 &, 0.0 &, 1.0)
\end{array} \\
\end{split}
\end{equation}
with $a=b$ (see Table~\ref{Tab_struc}) and $c=L$.
The atomic structures of the materials studied in this work are provided in Table~\ref{Tab_sublattice_coord}.
In the case of RhI$_{3}$, after having performed the computation with the structure described in Table~\ref{Tab_sublattice_coord}, we applied a counterclockwise rotation of $\pi/2$ to the calculated flexoelectric tensor, in order to recover the geometry shown in Fig.~2(e) of the main text.

\begin{table} 
\setlength{\tabcolsep}{8pt}
\begin{center}
\begin{tabular}{c|ccc} 
\hline \hline
 & \multicolumn{1}{c}{$a$ (\r{A})} & $b$ (\r{A})  & $h$(\r{A})  \\
 \hline
Si   & 3.814 & 3.814 & 0.215  \\
(blue) P    & 3.212 & 3.212 & 0.619  \\
BN   & 2.474 & 2.474 & 1.625   \\
SnS$_2$   & 3.618 & 3.618 & 1.468  \\
RhI$_3$ &6.671 & 6.671  & 1.488 \\
\hline \hline 
\end{tabular}
\end{center}
\caption{Equilibrium structural parameters for the unperturbed flat configuration of the materials considered in this work. $h$ corresponds to half the thickness of the buckled materials.}
\label{Tab_struc}
\end{table}


\begin{table} 
\setlength{\tabcolsep}{8pt}
\begin{center}
\begin{tabular}{c|ccc} 
\hline \hline
 & \multicolumn{1}{c}{$\tau_{1}$} & $\tau_{2}$  & $\tau_{3}$  \\
 \hline
Si$_{1}$   & $1/3$ & $2/3$ & $+h_{\rm{Si}}/L$  \\
Si$_{2}$  & $2/3$ & $1/3$ & $-h_{\rm{Si}}/L$  \\
\hline\hline
P$_{1}$  &  $1/3$ & $2/3$ & $+h_{\rm{P}}/L$  \\
P$_{2}$ & $2/3$ & $1/3$ & $-h_{\rm{P}}/L$  \\
\hline\hline
B$_{1}$   & $1/3$ & $2/3$ & $+h_{\rm{B}}/L$  \\
B$_{2}$   & $2/3$ & $1/3$ & $-h_{\rm{B}}/L$  \\
N$_{1}$   & $2/3$ & $1/3$ &  $+h_{\rm{N}}/L$ \\
N$_{2}$  & $1/3$ & $2/3$ &  $-h_{\rm{N}}/L$ \\
\hline\hline
Sn & $0$ & $0$ & $0$  \\
S$_{1}$   & $1/3$ & $2/3$ & $+h_{\rm{SnS}_{2}}/L$  \\
S$_{2}$   & $2/3$ & $1/3$ & $-h_{\rm{SnS}_{2}}/L$  \\
\hline\hline
Rh$_{1}$ & $1/6$ & $5/6$ & $0$  \\
Rh$_{2}$ & $5/6$ & $1/6$ & $0$  \\
I$_{1}$ & 0.148 & 0.5 &  $-h_{\rm{RhI}_{3}}/L$ \\
I$_{2}$ & 0.5 & 0.852 & $+h_{\rm{RhI}_{3}}/L$  \\
I$_{3}$ & 0.852 & 0.852 & $-h_{\rm{RhI}_{3}}/L$  \\
I$_{4}$ & 0.148  & 0.148 & $+h_{\rm{RhI}_{3}}/L$  \\
I$_{5}$ & 0.5 & 0.148 & $-h_{\rm{RhI}_{3}}/L$  \\
I$_{6}$ & 0.852 & 0.5 &  $+h_{\rm{RhI}_{3}}/L$ \\
\hline \hline 
\end{tabular}
\end{center}
\caption{Primitive-cell atomic structures, expressed in reduced coordinates, for the materials considered in this work. $h$ refers to the parameter reported in Table.\ref{Tab_struc}, except for BN-bilayer where B and N occupy sublattices positions described by slightly different $z$-components ($h_{B}=3.0661$ Bohr and $h_{N}=3.0703$ Bohr for B and N respectively).}
\label{Tab_sublattice_coord}
\end{table}

\section{Spontaneous axial polarization in nanotube structures}

\begin{table}[b!]
\setlength{\tabcolsep}{8pt}
\begin{center}
\begin{tabular}{c|cc|cc}\hline\hline
   R \T\B & \multicolumn{1}{c}{$R\cdot F^{\rm Sn}_z$} & \multicolumn{1}{r}{$R\cdot F^{\rm S}_z$} & $ P^{\rm 1D, CI}$ & $ P^{\rm 1D, LM}$  \\ \hline
  26.122 \T\B   & $-$0.135 & 0.066 & $-$0.800 & 1.106    \\
  39.183 \T\B   & $-$0.130 & 0.066 & $-$0.789 & 1.304    \\ \hline
  l.r. \T\B     & $-$0.132 & 0.065 & $-$0.790 & 1.422    \\
  \hline \hline
\end{tabular}
\caption{\label{Tab_NT} Results from direct calculations for nanotubes of two different radii $R$ [in bohr] (two top rows) and  from the linear-response calculation (bottom row). The two left-most columns show the average axial forces [times $R$ and in hartree units] on the Sn and S atoms. 
The two right-most columns show the clamped-ion and lattice-mediated 1D axial nanotube polarizations [in $e$]. Direct 1D polarizations are calculated as $\Omega P_z/c$, with $\Omega$, $c$ and $P_z$ being, respectively, the volume, dimension and 3D polarization component along the axial direction of the simulation supercell. In turn, linear response ones are calculated with Eq.~(9) of the main text.}
\end{center}
\end{table}

%
%
%
%
To test Eq.~(9) of the main text, 
we perform direct ground-state DFT calculations of zigzag SnS$_2$ nanotubes and extract the atomic
forces, together with the macroscopic electronic polarization along the axial direction via the Berry phase implementation 
of VASP.5.4~\cite{Kresse-93,Kresse-94,Kresse-96,Kresse-99}. 
We also perform a full structural relaxation of the nanotube structures to assess the LM contribution to $\mu$, and
hence to Eq.~(9) of the main text.

Each SnS$_{2}$ nanotube is placed in a simulation supercell that is hexagonal over the radial plane, with sufficient outer vacuum regions for the periodic replicas to be mechanically decoupled, and the tube axis is oriented along the $z$ Cartesian direction (with a lattice parameter $c=11.84$ bohr). 
We use the projector-augmented wave (PAW) approach with an energy cutoff of 520 eV, the local-density approximation (LDA) to the exchange and correlation potential, and we sample the first Brillouin zone with a grid of $1\times1\times10$ {\bf k} points centered at $\Gamma$. 
The atomic positions are relaxed until all the forces are smaller in magnitude than $2\times10^{-4}$ eV/\r{A}. 

In table~\ref{Tab_NT} we show the results obtained for two nanotubes generated by folding stripes of 24 and 36 SnS$_2$ 6-atom unit cells.
The CI electronic polarization and the atomic forces are in essentially perfect agreement with the linear-response values 
shown in the bottom row of the table.
Such a match is remarkable given the substantial differences in methods and codes between the two calculations.
The atomic relaxations show a slightly slower degree of convergence, which we ascribe to the frequency hardening 
of the contributing polar phonon with curvature (and therefore a physical effect, not a limitation of
our method).
Note the two orders of magnitude difference between the axial polarization of our SnS$_2$ nanotubes and the values 
reported for a bundle of very thin ($<7.5$ bohr radius) h-BN nanotubes.~\cite{nakhmanson2003spontaneous}


%
%
%

\section{Longitudinal electric fields induced by a flexural phonon}

Let us consider a flexural phonon traveling in a suspended quasi-2D crystal with a small in-plane momentum $\textbf{q} = (q_{x},q_{y},0)$.
In the first-order linear regime, the displacement pattern of such a mode can be written as a cell-periodic part times a phase as follows 
\begin{equation}
u^{l}_{\kappa \beta} = u_{\kappa\beta}^{\bf q} e^{\textbf{q}\cdot\textbf{R}_{l\kappa}} = U \left( \delta_{\beta z} - iq_{\beta}\tau_{\kappa z} \right) e^{\textbf{q}\cdot\textbf{R}_{l\kappa}},
\label{eq:distpat}
\end{equation}
where $\textbf{R}_{l\kappa}$ is the position of the atom $\kappa$ in the cell $l$, and we have performed a long-wavelength expansion of the cell-periodic part up to first order in the momentum (see e.g. Sec. 2.2 of the Supplemental Material of Ref.~\cite{springolo2021direct}).

We shall show here that the lattice distortion given by Eq.~\eqref{eq:distpat} induces long-range electrostatic
fields along the longitudinal (in-plane) direction.
To this end, we resort to the formalism developed in Refs.~\cite{Royo2021,ponce2022drift,ponce2022long}, where the in-plane electrostatic potential generated by a charge perturbation is expressed as
\begin{equation}
 V^{\parallel}(\textbf{q}) = 2\pi\dfrac{\rho^{\parallel}(\textbf{q}) }{q},
 \label{eq:evenpot}
\end{equation}
with $q=\vert\textbf{q}\vert$. The $\parallel$ superscript equivalently means that both the potential and charge density are even
functions with respect to a reflection along the out-of-plane direction.
The charge response to the perturbation parameter $U$ 
is built from the individual sublattices displacements as follows
\begin{equation}
\rho^\parallel({\bf q})  = \sum_{\kappa \beta} \left( \delta_{\beta z} - iq_{\beta}\tau_{\kappa z} \right) \rho_{\kappa \beta}^\parallel({\bf q}),
\end{equation}
with
\begin{equation}
\rho^{\parallel}_{\kappa\beta}(\textbf{q}) = \int dz \rho^{\textbf{q}}_{\kappa\beta}(z) \, \cosh(qz),
\end{equation}
and $\rho_{\kappa\beta}^{\textbf{q}}(z)$ being the in-plane average of the microscopic charge-density response to the displacement of atom $\kappa$ along $\beta$.

Our goal is to derive a practical formulation of Eq.~\eqref{eq:evenpot} by pursuing the usual long-wavelength expansion of the constituent objects.
This procedure requires to combine Eq.~\eqref{eq:distpat} with 
\begin{equation}
\cosh(qz) = 1 + \frac{q^2 z^2}{2} + \cdots,
\end{equation}
and with the long-wavelength expansion of the microscopic charge-density respsonse
\begin{equation}
\rho_{\kappa \beta}^{\bf q} (z) = \rho_{\kappa \beta} (z) - iq_\gamma \rho_{\kappa \beta}^{(1,\gamma)} (z)
    - \frac{q_\gamma q_\delta}{2} \rho_{\kappa \beta}^{(2,\gamma \delta)} (z) + i \frac{q_\gamma q_\delta q_\lambda}{6}\rho_{\kappa \beta}^{(3,\gamma \delta \lambda)} (z) \cdots
\end{equation}
to arrive at a corresponding expansion for $\rho^\parallel({\bf q})$. At zero-th and first order in ${\bf q}$, 
$\rho^\parallel({\bf q})$ vanishes because of the acoustic sum rule and charge neutrality; at second order it
also vanishes because of rotational invariance. 
We therefore focus on the leading $O(q^3)$ contributions to $\rho^\parallel({\bf q})$, 
\begin{equation}
\begin{split}
\rho^\parallel ({\bf q})  = & - i\frac{q^2 q_{\beta}}{2} \sum_{\kappa} \Bigg[ 
 \int dz \, z^2 \rho_{\kappa z}^{(1,\beta)} (z) + \tau_{\kappa z}  \int dz \, z^2 \rho_{\kappa \beta} (z) \Bigg] \\
  & + i \frac{q_\beta q_\delta q_\lambda}{6} \sum_{\kappa} \Bigg[ \int dz \,  \rho_{\kappa z}^{(3,\beta \delta\lambda)} (z) 
   + 3 \tau_{\kappa z} \int dz \,  \rho_{\kappa \beta}^{(2,\delta\lambda)} (z) \Bigg] + O(q^4).
  \label{eq:rhopoles}
\end{split}
\end{equation}
The first line on the rhs vanishes identically. To prove this, 
we use the following relations~\cite{artlin},
\begin{equation}
\sum_\kappa \rho_{\kappa z}^{(1,\gamma)} ({\bf r}) = \sum_\kappa \rho_{\kappa \gamma}^{(1, z)} ({\bf r}) 
     = \sum_\kappa (z - \tau_{\kappa z}) \rho_{\kappa \gamma}({\bf r}).
\end{equation}
Then the sublattice sum of the first square bracket reduces to
\begin{equation}
\int dz \, z^3 \sum_{\kappa} \rho_{\kappa \gamma} (z) = - \dfrac{1}{S}\int d^{3}r \, z^3 \dfrac{\partial \rho(\textbf{r})}{\partial r_{\gamma}}
\label{tr_inv}
\end{equation}
(where translational invariance has been used in the last equality and $\rho(\textbf{r})$ corresponds the microscopic charge density within the undistorted configuration),
which is manifestly zero after integration by parts (recall that $\gamma$ here refers to an in-plane direction). 
In the end, we reach a simpler final expression that can be written as follows (summation over repeated indices is implied)
\begin{equation}
\rho^\parallel ({\bf q})  = i q_\alpha q_\beta q_\gamma  \left(
  \frac{1}{6S} \sum_\kappa {O}^{(\alpha\beta\gamma)}_{\kappa z} +  \frac{1}{2S} \sum_\kappa \tau_{\kappa z} {Q}_{\kappa \alpha}^{( \beta\gamma)}\right),
\label{2D_octupole}
\end{equation}
where $S$ is the in-plane area of the primitive cell while ${\bf Q}$ and ${\bf O}$ are, respectively the 3D 
dynamic quadrupole and octupole tensors of the supercell.
This result can be further simplified after recalling the following relations between charge density and polarization moments~\cite{artlin}
\begin{align}
\dfrac{1}{2}\sum_{k} O^{(\alpha\beta\gamma)}_{\kappa z} = & \Omega \left(\bar{\mu}^{\rm{I}}_{\alpha z,\beta\gamma} + \bar{\mu}^{\rm{I}}_{\beta z,\gamma\alpha} + \bar{\mu}^{\rm{I}}_{\gamma z,\alpha\beta} \right), \\
Q^{(\alpha\beta)}_{\kappa \gamma} = & \Omega \left( \bar{P}^{(1,\beta)}_{\alpha,\kappa\gamma} + \bar{P}^{(1,\alpha)}_{\beta,\kappa\gamma} \right).
\end{align}
After gathering all terms in Eq.(\ref{2D_octupole}) (and using the relation $\Gamma^{\kappa}_{\alpha ,z\beta} = \Gamma^{\kappa}_{\alpha ,\beta z} = -\delta_{\alpha\beta} \tau_{\kappa z}$ valid for an isolated slab), we find
\begin{equation}
\rho^{\parallel}(\textbf{q}) = i q_{\alpha}q_{\beta}q_{\gamma} \dfrac{L}{3} \left( \mu^{\rm{I},\rm{el}}_{\alpha z,\beta\gamma} + \mu^{\rm{I},\rm{el}}_{\beta z,\gamma\alpha} + \mu^{\rm{I},\rm{el}}_{\gamma z,\alpha\beta} \right) + O(q^{4})
\end{equation}
Recalling the relation proved in section \ref{mu2D_muI}
\begin{equation}
\mu^{2\rm{D},\rm{CI}}_{\alpha z,\beta\gamma} = -L \mu^{\rm{I},\rm{el}}_{\alpha z,\beta\gamma}
\end{equation}
we arrive at
\begin{equation}
\label{rhopar}
\rho^{\parallel}(\textbf{q}) = -i q_{\alpha}q_{\beta}q_{\gamma} \mu^{2\rm{D},\rm{CI}}_{\alpha z,\beta\gamma}  + O(q^{4}).
\end{equation} 
This result, which can be trivially generalized to the relaxed-ion case, in combination with Eq.~(\ref{eq:evenpot}) 
yields Eq.~(10) of the main text. 

Note that, in the flexural phonon considered in this Section, the symmetrized strain-gradient tensor reads as
\begin{equation}
\varepsilon_{\beta \gamma,z}({\bf r}) = U q_\beta q_\gamma e^{i {\bf q\cdot r}},
\end{equation}
and via Eq.~(2) of the main text yields a flexoelectric polarization 
\begin{equation}
\frac{\partial P_\alpha ({\bf r})}{\partial U} =  q_\beta q_\gamma \mu^{2\rm{D},\rm{CI}}_{\alpha z,\beta\gamma} e^{i {\bf q\cdot r}}.
\end{equation}
By applying $\rho = -\nabla \cdot {\bf P}$, we trivially recover Eq.~(\ref{rhopar}), proving
the consistency of our definition of $\bm{\mu}^{\rm 2D}$ with the fundamental treatment of
electrostatics in 2D developed in Ref.~\cite{Royo2021}.

\section{Choice of the exchange-correlation functional}

\begin{table} [b!]
\setlength{\tabcolsep}{6pt}
\begin{center}

\begin{tabular}{c|r|r|r|r}\hline\hline
    \T\B & \multicolumn{1}{c}{} & \multicolumn{1}{c}{$\mu^{2\rm{D}}_{yz,xx}$} & \multicolumn{1}{c}{} & \multicolumn{1}{|c}{$\mu^{2\rm{D}}_{zz,xx}$} \\ \hline
    \T\B & \multicolumn{1}{c}{CI} & \multicolumn{1}{c}{LM} & \multicolumn{1}{c}{RI} & \multicolumn{1}{|c}{RI} \\ \hline
   Si \T\B  & 0.0239  & 0.0000 & 0.0239 & 0.0016 \\
        P \T\B   & 0.0669  & 0.0000  & 0.0669 & 0.0206 \\
        SnS$_{2}$ \T\B & 0.1369  &  $-$0.2016 & $-$0.0646 & 0.0178 \\
        RhI$_{3}$ \T\B & $-$0.1733 & 0.0202 & $-$0.1530 & $-$0.0080  \\
 2-BN (direct) \T\B  & 1.0013 & $-$0.6215 & 0.3797 & $-$0.0254  \\
 2-BN (model) \T\B  & 0.9986  & $-$0.6213  & 0.3773  &   \\
  \hline \hline
\end{tabular}
\caption{Calculated 2D flexoelectric coefficients within PBE.
Left columns show the Clamped-Ion(CI), Lattice-Mediated(LM), and Relaxed-Ion(RI) contributions to the in-plane response $\mu^{2\rm{D}}_{yz,xx}$. The right column shows the out-of-plane RI response, $\mu^{2\rm{D}}_{zz,xx}$. Results are provided in units of electronic charge. 2-BN corresponds to the BN bilayer, which we calculated either directly or by means of the piezoelectric model ($\mu^{2\rm{D}}= E h$) within PBE. The values of the CI and LM longitudinal piezoelectric constant are $E^{\rm{CI}}=0.1208$ e/bohr and $E^{\rm{LM}}=-0.0749$ e/bohr, respectively, in good agreement with the LDA values reported in the main text. The calculated PBE interlayer distance is $h = 8.29$ bohr.}
\label{GGA_mu}
\end{center}
\end{table}

To verify the impact of the exchange-correlation functional, we recalculated the flexoelectric coefficient 
$\mu^{2\rm{D}}$ defined in the main text by using the PBE functional in place of LDA; the results are presented in 
Table \ref{GGA_mu}.
The LDA and PBE results show reasonable agreement in most cases, 
with typical deviations that are in line with the expectations (e.g., similar
deviations were pointed out for the out-of-plane response in our earlier work~\cite{springolo2021direct}).
The largest disagreement occurs in the case of the BN bilayer, 
with a PBE flexoelectric coefficient that is 35\% larger than the LDA value.
Interestingly, a closer look at the relaxed PBE structure reveals that the equilibrium 
interlayer distance, $h_{\rm{PBE}} = 8.29$ bohr, displays a comparable (35\%) overestimation 
respect to the LDA result ($h_{\rm{LDA}} = 6.14$ bohr).
%
Given that the noninteracting-layer formula, $\mu=Eh$, accurately holds in both
cases (compare the ``direct'' and ``model'' row in Table~\ref{GGA_mu}), 
the large disagreement in the interlayer distance $h$ is almost entirely responsible for 
the discrepancy in the calculated $\mu$.
(The piezoelectric coefficient of monolayer BN, $E$, has similar values in LDA and PBE.)

Such a discrepancy in the value of $h$ is not surprising, and arises from the fact that a bilayer system is a 
van der Waals bonded compound, i.e., a classic situation where local and semilocal approximations to DFT fail.
PBE, in particular, does not seem to bind the two BN layers at all, which explains the unusually large 
value of $h$. 
This structural parameter has been calculated by taking van der Waals corrections into
account, at various levels of theory, in~\cite{PhysRevB.92.155438}.
The most reliable values quoted therein, of $h=3.34-3.51$ \AA, are similar (3--9\% larger) to our LDA 
value, and 20\% smaller than the PBE one.
All in all, this analysis indicates that whenever the vdW corrections are needed, the flexoelectric response
of the system is trivially given as a weighted sum of the piezoelectric response of the constituents,
in agreement with the conclusions of \cite{duerloo2013flexural}.


%
%

%

%

\begin{figure}[b!] 
\begin{center}
  \includegraphics[width=0.9\textwidth]{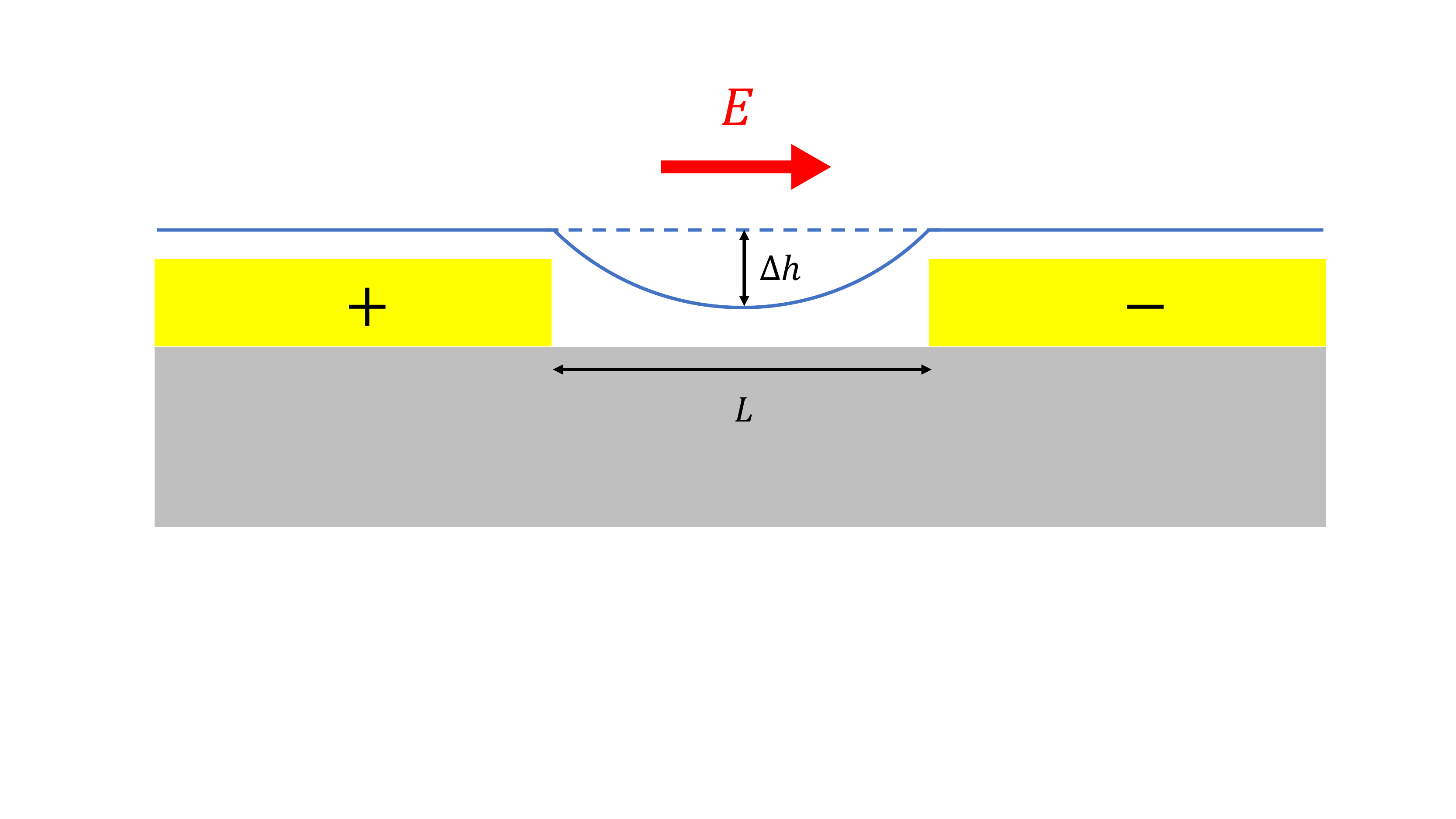}\\
  \caption{Schematic representation of the experimental setup proposed. The layer is deposited above two electrodes separated by a distance $L$ and in turns leaned on a substrate. The two electrodes are used to apply an in-plane uniform electric field $E$ inducing a curvature of the layer in corrispondence with the hole, where the layer is allowed to deform.}
\label{Fig_setup}
\end{center}
\end{figure}

\section{Experimental setup}

In this section we propose two possible experimental setups that could be used to confirm our theoretical predictions. 

\subsection{External electric field-induced bending}

Let us consider the system schematically illustrated in Fig.\ref{Fig_setup}. 
It consists of a single layer (blue) deposited on a gap between two metallic electrodes (yellow),
both lying on a flat substrate (grey). The system can be regarded as extended along 
the normal direction (not shown).
Depositing 2D layers on nanopatterned substrates with gaps of various shapes and dimensions 
is a well established experimental practice, see e.g. Ref.~\cite{Steeneken_2021}, so the setup of Fig.\ref{Fig_setup}
should be reasonably straightforward to prepare.

%
%

By applying an alternating in-plane electric field, our theory predicts that the layer-membrane will start to 
vibrate vertically at the same frequency, at an amplitude that depends on the crystallographic orientation of
the layer. In particular, the response should be largest when one of the mirror planes of the crystal is parallel 
to the direction of the electric field, and vanish if one of such planes is perpendicular to it.
By probing the system near the electromechanical resonance (within the linear regime, the suspended layer 
can be accurately modeled as a damped harmonic oscillator~\cite{Steeneken_2021}), a detectable feature should appear 
in the measured capacitance.
The amplitude of such signal, together with the relevant mechanical and geometrical parameters of the 
system, should provide a reliable quantitative estimatimate of the flexoelectric coefficient that we define
and calculate in this work.

As an alternative, one could also probe the vertical displacements of the layer directly, e.g., via an AFM tip.
In such case, the electromechanical response of the system might be complicated by the coupling to the tip, but
the most important qualitative signatures of the effect (e.g. its directional dependence) should be detectable 
nonetheless.

\subsection{Second harmonic generation method}

The second experimental setup builds on the use of the Second Harmonic Generation (SHG) technique.
As we illustrate in the main text, rippled and bent geometries are decorated 
with topologically nontrivial polarization textures via the physical effect we discuss here.
SHG is sensitive to a local breakdown of space inversion (SI) symmetry and on the direction
along which SI is broken; it could be used, therefore, to image the predicted polarization patterns 
directly.
The use of SHG in 2D systems is well documented (see, e.g.,~\cite{Zhang_2020}), and 
should therefore be applicable to characterizing the effects that we describe.

As a means to obtain the desired ripple patterns, one could again use nanopatterning techniques,
as explained in the above paragraphs. Indeed, suspended layers are seldom flat: they typically feel 
the van der Waals attraction of the far-away substrate and tend to bend inward. This
means that an in-plane polarization should be present in a suspended layer, with a 
topology that depends on the shape of the gap, even in absence of an applied electric field.

\section{Topological polarization textures in arbitrary rippled structures}

For a generic ripple $u_{z}(x,y)$, at the first order in the local shape-operator tensor field $b_{\beta\gamma}(x,y)$, the induced 2D local polarization vector field can be written as
\begin{equation}
P^{2\rm{D}}_{\alpha}(x,y) = \mu^{2\rm{D}}_{\alpha z,\beta\gamma} b_{\beta\gamma}(x,y),
\end{equation}
with $\beta,\gamma$ referring to in-plane directions while $\alpha = x,y,z$.
$\mu^{2\rm{D}}$ is the 2D flexoelectric coefficient as defined in the main text, for $\alpha= x,y$, and as $\mu^{2\rm{D}} = \epsilon_{0}\varphi$, for $\alpha=z$~\cite{springolo2021direct}.
Sum over repeated indices is understood.
In turn, the local shape-operator tensor field is given by
\begin{equation}
b_{\beta\gamma}(x,y) = \dfrac{\partial^{2}u_{z}(x,y)}{\partial r_{\beta}r_{\gamma}} .
\end{equation}

Within the symmetries of the materials considered, we have
\begin{equation}
\begin{split}
& P^{2\rm{D}}_{x}(x,y) = 2 \mu^{2\rm{D}} b_{xy}(x,y), \\
& P^{2\rm{D}}_{y}(x,y) =  \mu^{2\rm{D}} \left( b_{xx}(x,y) - b_{yy}(x,y) \right), \\
& P^{2\rm{D}}_{z}(x,y) = \epsilon_{0}\varphi Tr(b),
\end{split} 
\end{equation}
where $Tr(b) = \sum_{\beta} b_{\beta\beta}(x,y)$ is the trace of the local shape-operator field.

\begin{figure}
\centering
\includegraphics[height=2.0in]{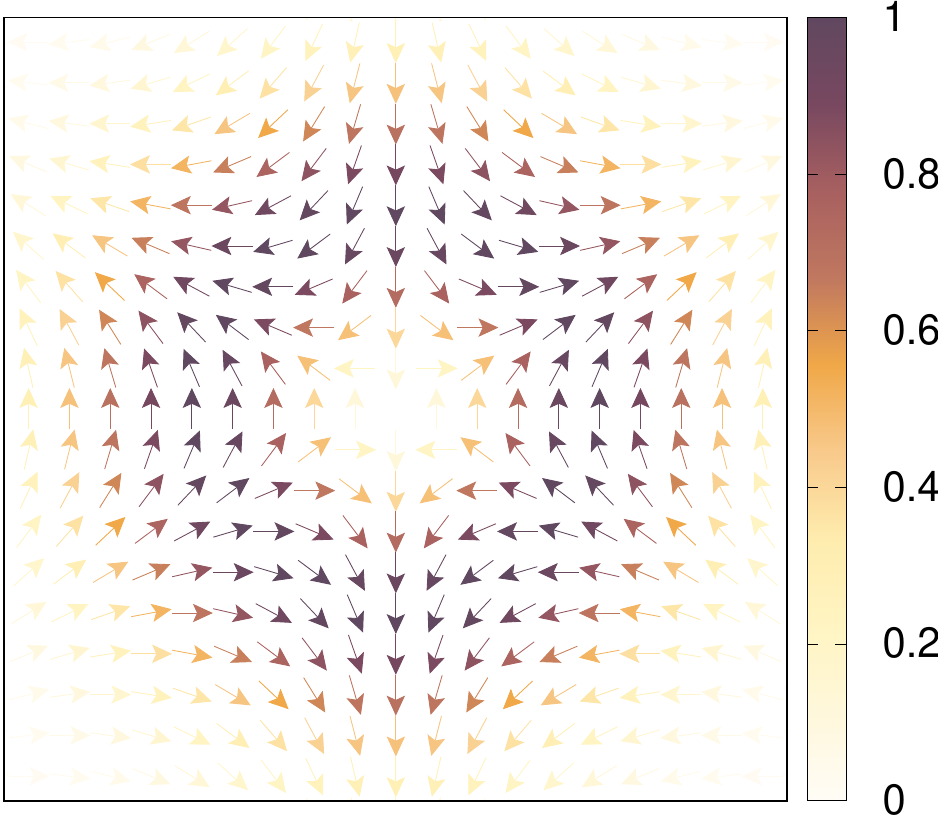}
\hspace{0.5cm}
\includegraphics[height=2.0in]{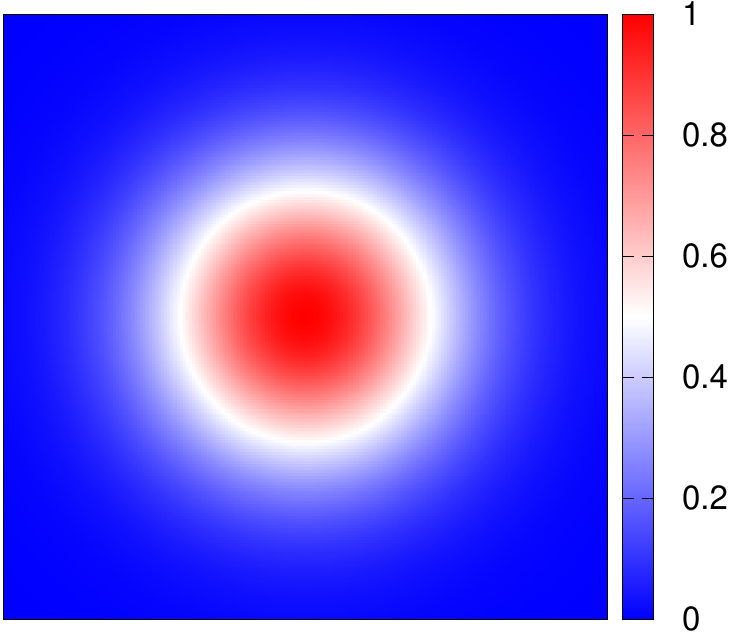}
\includegraphics[height=2.0in]{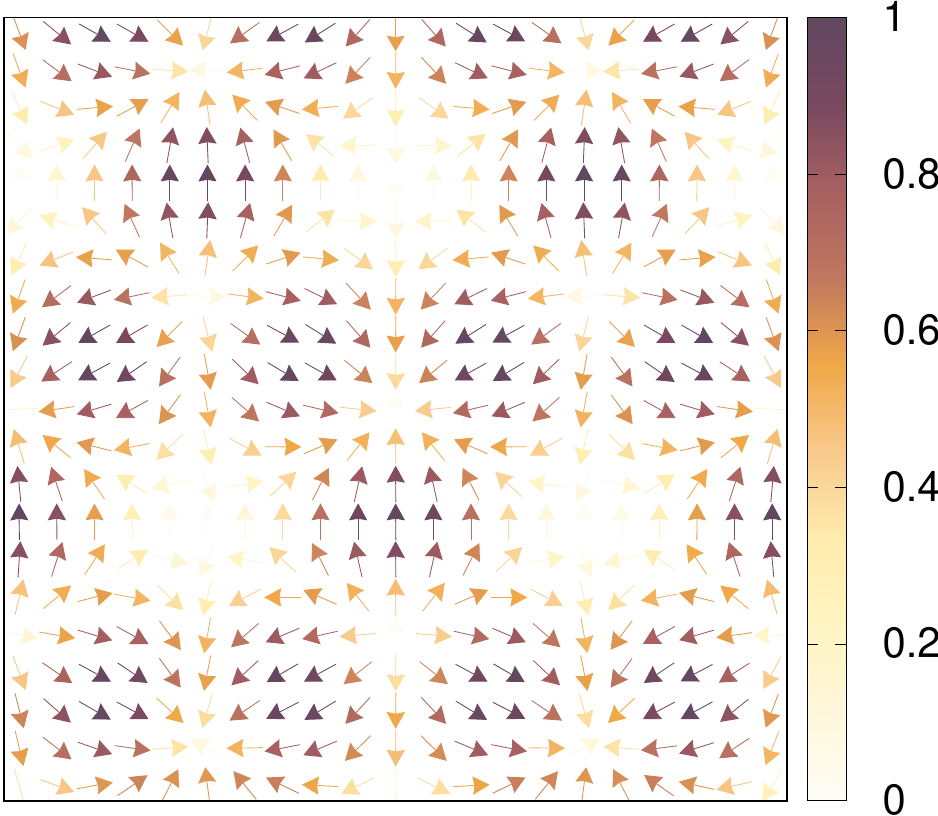}
\hspace{0.5cm}
\includegraphics[height=2.0in]{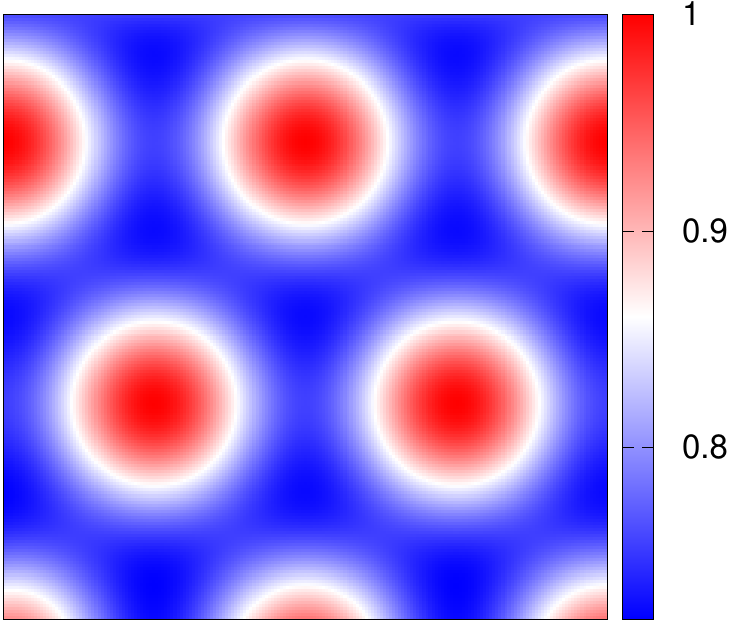}
\caption{Top and bottom left panels: Polarization texture associated with a single Gaussian bump of the type $z=Ae^{-(x^2+y^2)/\sigma^2}$ and an hexagonal lattice of Gaussian bumps of the same type. The arrows indicate the polarization direction, its amplitude (in units of $\vert P_{\parallel} \vert^{\rm{max}} \simeq 1.48 \dfrac{A\mu}{\sigma^{2}}$ and $\vert P_{\parallel} \vert^{\rm{max}} \simeq 6.94 \dfrac{A\mu}{\sigma^{2}}$ rispectively) is defined by the color scale.  Top and bottom right panels: Contour plot for the considered deformation. Their amplitudes (in units of $u^{\rm{max}}_{z}=A$ and $u^{\rm{max}}_{z}\simeq 1.11 A$ respectively) are defined by the color scale. }
\label{Q_m2}
\end{figure}

\begin{figure}
\centering
\includegraphics[height=2.0in]{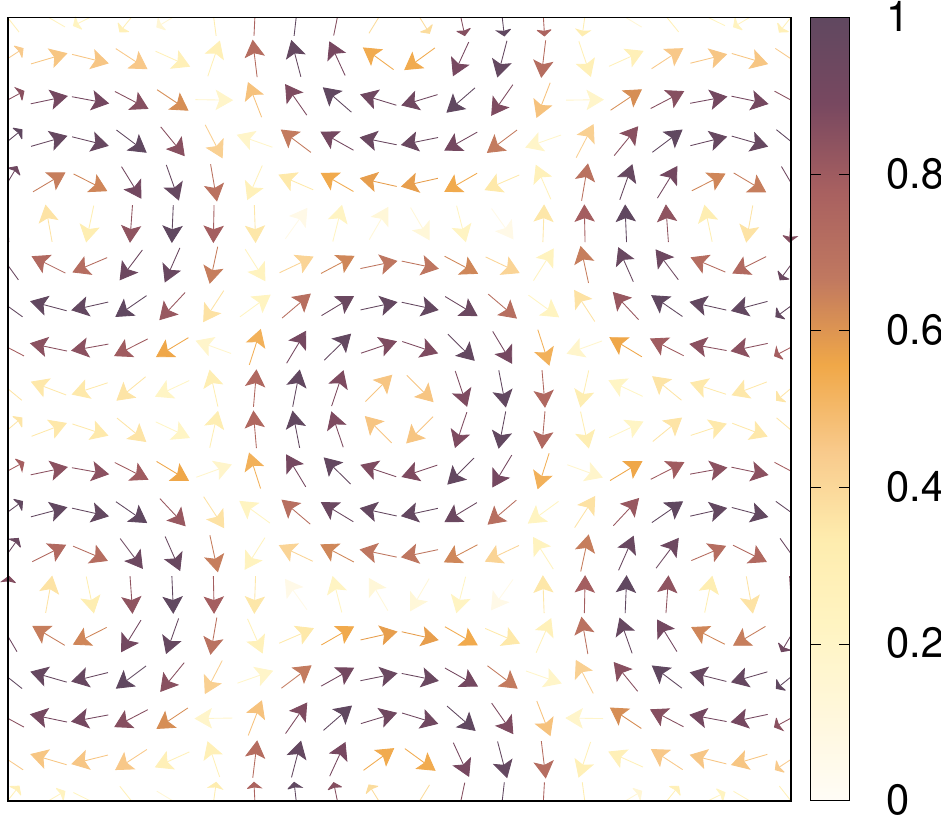}
\hspace{0.5cm}
\includegraphics[height=2.0in]{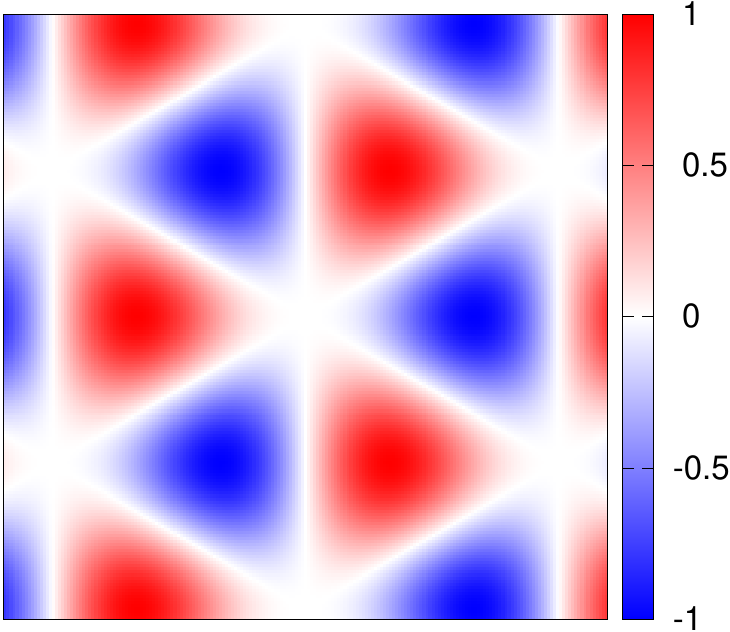}
\includegraphics[height=2.0in]{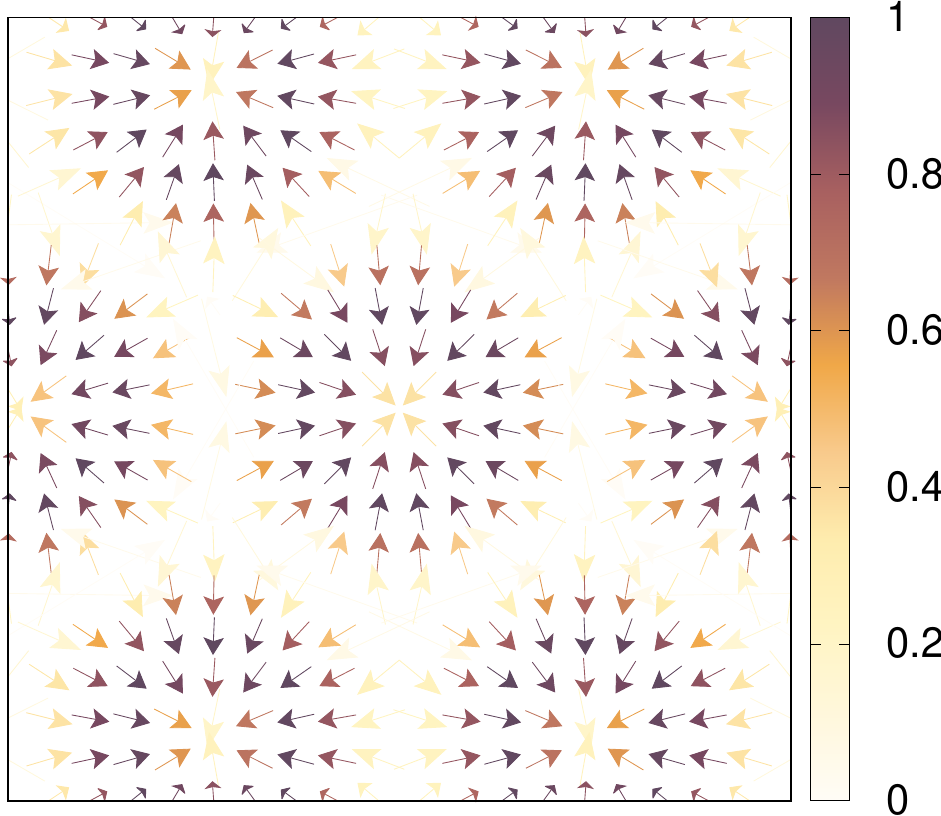}
\hspace{0.5cm}
\includegraphics[height=2.0in]{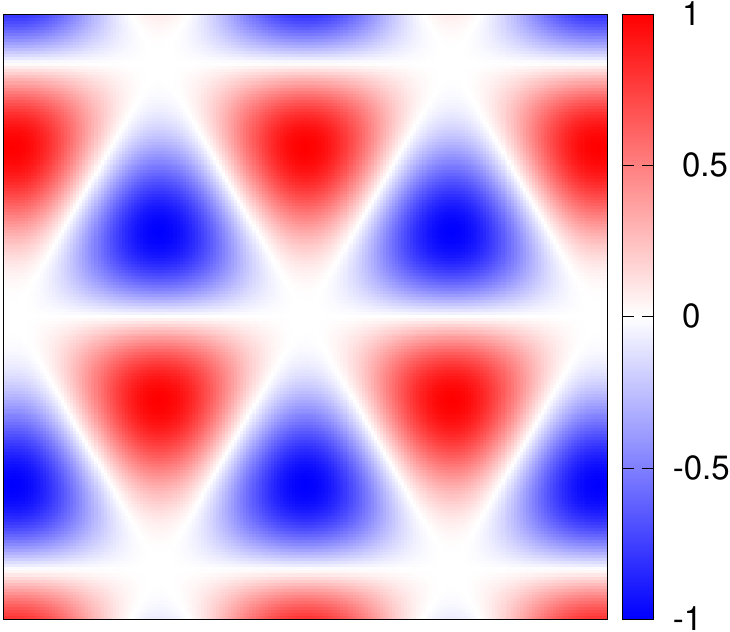}
\caption{Top and bottom left panels: Polarization texture associated with two periodic patterns of the type $z=A\sum_{i=1}^3 \sin({\bf q_i}\cdot {\bf r})$, with ${\bf q}_1 = q(1,0,0)$, ${\bf q}_2 = q(-1/2,\sqrt{3}/2,0)$, ${\bf q}_3 = q(-1/2,-\sqrt{3}/2,0)$, and ${\bf q}_1 = q(\sqrt{3}/2,1/2,0)$, ${\bf q}_2 = q(-\sqrt{3}/2,1/2,0)$, ${\bf q}_3 = q(0,-1,0)$ respectively.  $q=2\pi/L$. The arrows indicate the polarization direction, its amplitude (in units of $\vert P_{\parallel} \vert^{\rm{max}} =1.75 A\mu q^2$) is defined by the color scale.  Top and bottom right panels: Contour plot of the periodic deformations considered. Their amplitudes (in units of $u^{\rm{max}}_{z}\simeq 2.6 A$) are defined by the color scale. }
\label{Q_1}
\end{figure}

In Fig.\ref{Q_m2} and \ref{Q_1} we report the resulting in-plane polarization field associated to several types of ripples, within the kind of materials considered in the main text.


\bibliography{refs1}
\bibliographystyle{styletesis}